\documentclass[a4paper,usenatbib,times,fleqn]{mnras}
\usepackage{siunitx}
\usepackage{natbib,graphicx,amsmath,amssymb}
\usepackage{aas_macros,url,bm,float,placeins}
\usepackage[capitalise]{cleveref}
	\crefname{equation}{Equation}{Equations}
	\crefname{figure}{Figure}{Figures}
	\crefname{table}{Table}{Tables}
	
\usepackage{flushend}
\usepackage[usenames, dvipsnames]{color}
\usepackage{booktabs}

\newcommand{\mrm}[1]{_\mathrm{#1}}
\newcommand{\V}{\mathrm{\textbf{V}}}
\newcommand{\vecuv}{\mathrm{\textbf{v}}}
\newcommand{\veck}{\mathrm{\textbf{k}}}
\newcommand{\rhop}{\rho\mrm{p}}
\newcommand{\rhog}{\rho\mrm{g}}
\newcommand{\dt}[1]{\dfrac{\partial #1}{\partial t}}

\newcommand{\bnabla}{\boldsymbol{\nabla}}

\newcommand{\tstop}{t\mrm{stop}}
\newcommand{\OK}{\Omega\mrm{K}}

\newcommand{\vecu}{\mathrm{\textbf{u}}}
\newcommand{\ol}[1]{\overline{#1}}

\newcommand{\delg}{\delta_\mathrm{g}}
\newcommand{\delp}{\delta_\mathrm{p}}

\graphicspath{{figures/}}

\title[Streaming instability in pressure bumps]{Linear growth of streaming instability in pressure bumps}

\author[Auffinger \& Laibe]{J\'er\'emy Auffinger$^{1}$, Guillaume Laibe$^{1}$\thanks{E-mail: \href{mailto:guillaume.laibe@ens-lyon.fr}{mailto:guillaume.laibe@ens-lyon.fr}},  \\
$^{1}$Univ Lyon, Univ Lyon1, Ens de Lyon, CNRS, Centre de Recherche Astrophysique de Lyon UMR5574, F-69230, Saint-Genis-Laval, France
}
\pagerange{\pageref{firstpage}--\pageref{lastpage}} \pubyear{2017}

\begin{document}
%
%  These Macros are taken from the AAS TeX macro package version 4.0.
%  Include this file in your LaTeX source only if you are not using
%  the AAS TeX macro package and need to resolve the macro definitions
%  in the BibTeX entries returned by the ADS abstract service.
%
%  For more information on the AASTeX macro package, please see the URL
%	http://www.aas.org/publications/aastex.html
%  For more information about ADS abstract server, please see the URL
%	http://adswww.harvard.edu/ads_abstracts.html
%

% Abbreviations for journals.  The object here is to provide authors
% with convenient shorthands for the most "popular" (often-cited)
% journals; the author can use these markup tags without being concerned
% about the exact form of the journal abbreviation, or its formatting.
% It is up to the keeper of the macros to make sure the macros expand
% to the proper text.  If macro package writers agree to all use the
% same TeX command name, authors only have to remember one thing, and
% the style file will take care of editorial preferences.  This also
% applies when a single journal decides to revamp its abbreviating
% scheme, as happened with the ApJ (Abt 1991).

\def\jnl@style{\it}
%commente par Seb
\def\aaref@jnl#1{{\jnl@style#1}}
%ref remplace par aaref pour eviter conflit...

\def\aaref@jnl#1{{\jnl@style#1}}

\def\aj{\aaref@jnl{AJ}}                   % Astronomical Journal
\def\araa{\aaref@jnl{ARA\&A}}             % Annual Review of Astron and Astrophys
\def\apj{\aaref@jnl{ApJ}}                 % Astrophysical Journal
\def\apjl{\aaref@jnl{ApJ}}                % Astrophysical Journal, Letters
\def\apjs{\aaref@jnl{ApJS}}               % Astrophysical Journal, Supplement
\def\ao{\aaref@jnl{Appl.~Opt.}}           % Applied Optics
\def\apss{\aaref@jnl{Ap\&SS}}             % Astrophysics and Space Science
\def\aap{\aaref@jnl{A\&A}}                % Astronomy and Astrophysics
\def\aapr{\aaref@jnl{A\&A~Rev.}}          % Astronomy and Astrophysics Reviews
\def\aaps{\aaref@jnl{A\&AS}}              % Astronomy and Astrophysics, Supplement
\def\azh{\aaref@jnl{AZh}}                 % Astronomicheskii Zhurnal
\def\baas{\aaref@jnl{BAAS}}               % Bulletin of the AAS
\def\jrasc{\aaref@jnl{JRASC}}             % Journal of the RAS of Canada
\def\memras{\aaref@jnl{MmRAS}}            % Memoirs of the RAS
\def\mnras{\aaref@jnl{MNRAS}}             % Monthly Notices of the RAS
\def\pra{\aaref@jnl{Phys.~Rev.~A}}        % Physical Review A: General Physics
\def\prb{\aaref@jnl{Phys.~Rev.~B}}        % Physical Review B: Solid State
\def\prc{\aaref@jnl{Phys.~Rev.~C}}        % Physical Review C
\def\prd{\aaref@jnl{Phys.~Rev.~D}}        % Physical Review D
\def\pre{\aaref@jnl{Phys.~Rev.~E}}        % Physical Review E
\def\prl{\aaref@jnl{Phys.~Rev.~Lett.}}    % Physical Review Letters
\def\pasp{\aaref@jnl{PASP}}               % Publications of the ASP
\def\pasj{\aaref@jnl{PASJ}}               % Publications of the ASJ
\def\qjras{\aaref@jnl{QJRAS}}             % Quarterly Journal of the RAS
\def\skytel{\aaref@jnl{S\&T}}             % Sky and Telescope
\def\solphys{\aaref@jnl{Sol.~Phys.}}      % Solar Physics
\def\sovast{\aaref@jnl{Soviet~Ast.}}      % Soviet Astronomy
\def\ssr{\aaref@jnl{Space~Sci.~Rev.}}     % Space Science Reviews
\def\zap{\aaref@jnl{ZAp}}                 % Zeitschrift fuer Astrophysik
\def\nat{\aaref@jnl{Nature}}              % Nature
\def\iaucirc{\aaref@jnl{IAU~Circ.}}       % IAU Cirulars
\def\aplett{\aaref@jnl{Astrophys.~Lett.}} % Astrophysics Letters
\def\apspr{\aaref@jnl{Astrophys.~Space~Phys.~Res.}}
                % Astrophysics Space Physics Research
\def\bain{\aaref@jnl{Bull.~Astron.~Inst.~Netherlands}} 
                % Bulletin Astronomical Institute of the Netherlands
\def\fcp{\aaref@jnl{Fund.~Cosmic~Phys.}}  % Fundamental Cosmic Physics
\def\gca{\aaref@jnl{Geochim.~Cosmochim.~Acta}}   % Geochimica Cosmochimica Acta
\def\grl{\aaref@jnl{Geophys.~Res.~Lett.}} % Geophysics Research Letters
\def\jcp{\aaref@jnl{J.~Chem.~Phys.}}      % Journal of Chemical Physics
\def\jgr{\aaref@jnl{J.~Geophys.~Res.}}    % Journal of Geophysics Research
\def\jqsrt{\aaref@jnl{J.~Quant.~Spec.~Radiat.~Transf.}}
                % Journal of Quantitiative Spectroscopy and Radiative Transfer
\def\memsai{\aaref@jnl{Mem.~Soc.~Astron.~Italiana}}
                % Mem. Societa Astronomica Italiana
\def\nphysa{\aaref@jnl{Nucl.~Phys.~A}}   % Nuclear Physics A
\def\physrep{\aaref@jnl{Phys.~Rep.}}   % Physics Reports
\def\physscr{\aaref@jnl{Phys.~Scr}}   % Physica Scripta
\def\planss{\aaref@jnl{Planet.~Space~Sci.}}   % Planetary Space Science
\def\procspie{\aaref@jnl{Proc.~SPIE}}   % Proceedings of the SPIE

\let\astap=\aap
\let\apjlett=\apjl
\let\apjsupp=\apjs
\let\applopt=\ao

\label{firstpage}
\bibliographystyle{mnras}
\maketitle

\begin{abstract}
Streaming instability is a powerful mechanism which concentrates dust grains in protoplanetary discs, eventually up to the stage where they collapse gravitationally and form planetesimals. Previous studies inferred that it should be ineffective in viscous discs, too efficient in inviscid discs, and may not operate in local pressure maxima where solids accumulate. From a linear analysis of stability, we show that streaming instability behaves differently inside local pressure maxima. Under the action of the strong differential advection imposed by the bump, a novel unstable mode develops and grows even when gas viscosity is large. Hence, pressure bumps are found to be the only places where streaming instability occurs in viscous discs. This offers a promising way to conciliate models of planet formation with recent observations of young discs. 
\end{abstract}


 %=======================================================================================================

\section{Introduction}
\label{sec:intro}

The main challenge of planet formation consists in figuring out how solids originating from the interstellar medium concentrate and grow over orders of magnitude up to form planetary cores \citep{Chiang2010}. Up to decimetric sizes, surface forces are strong enough for dust grains to grow by hit-and-stick collisions \citep{Blum2008}. This is not the case anymore for larger pebbles, and solid aggregates are expected instead to undergo bouncing or fragmentation (e.g. \citealt{Guttler2010,Zsom2010}). On the other hand, rocky structures should typically reach hundreds of metres in size to be glued by their own gravity. Thus, a third mechanism must bridge the gap and collect pebbles up the the stage where their local weight becomes sufficient. \citet{Goodman2000} suggested that such a concentration may originate from an hydrodynamical instability. \citet{Youdin2005,Youdin2007} demonstrated that the flow resulting from the radial drift of dust particles in weakly viscous discs is actually linearly unstable. In the non-linear regime, this so-called streaming instability develops dust over-concentrations \citep{JS2007}, which may ultimately form planetesimals by gravitational instabilities in discs of sufficient metallicities \citep{Johansen2007,Johansen2009,Bai2010b,Bai2010c,Carrera2015}. Hence, streaming instability may be responsible for the Initial Mass Function of planetesimals in discs \citep{Simon2016,Schafer2017}. The robustness of the streaming instability has been tested against several numerical schemes \citep{Balsara2009,Miniati2010,Tilley2010,Bai2010,Johansen2012,Johansen2014}, towards the aim of simulating its effect in a global disc \citep{Lyra2013,Kowalik2013,Yang2014}. Other physical processes such as vortices \citep{Raettig2015}, photo-evaporation \citep{Carrera2017}, presence of small grains \citep{Laibe2014}, grain growth \citep{Dra2014} or snow lines \citep{Schoonenberg2017} may reinforce the ability of streaming instability to concentrate dust. 

Discs may therefore contain only a moderate amount of dust grains, since up to $50 \%$ of their retained solid material may be converted into planetesimals (e.g. \citealt{Johansen2015,Dra2016}). However, the emission from a continuous dense phase of millimetre grains is commonly detected in young discs (e.g. \citealt{HLTau2015,Andrews2016}), except at some specific locations. Dark rings are often associated to ongoing planet formation (e.g. \citealt{Zhang2015,Gonzalez2015,Okuzumi2016}) or even to planets (e.g. \citealt{Dipierro2015,Dong2015,Picogna2015,Rosotti2016}). To explain the persistence of the dust population almost everywhere, one may invoke the turbulent viscosity of the gas, which damps efficiently the small-scale perturbations at which streaming instability develops, but this would prevent planetesimal formation (see however \citealt{Johansen2007,Dittrich2013}). 

Such an interpretation is based on properties of the streaming instability derived for discs with monotonically decreasing pressure profiles. On the other hand, local pressure maxima may be created in the disc at some locations by internal processes (e.g. \citealt{Ruge2016,Estrada2016,Bethune2016,GLM2017}). These pressure bumps are privileged locations for planetesimal formation since they concentrate dust (e.g. \citealt{NSH1986,Haghighipour2005}). Finding a way for streaming instability to develop specifically in these pressure bumps would conciliate current scenarii of planetesimal formation and recent observations of discs. Simulations have been recently performed to show that streaming instability may deform the bump \citep{Taki2016}. So far, the resilience of the streaming instability against viscosity in pressure bumps has not been investigated.

In this study, we show that the development of the streaming instability in local pressure maxima is more complex than for discs with monotonic pressure profiles. We address the problem analytically by performing a linear perturbation analysis in a shearing box centred around a pressure maximum. In Sect.~\ref{sec:motion}, we derive new solutions for the steady state, since pressure curvature provides additional advection compared to the usual case. The analysis of the unstable modes of the system as a function of the steepness of the bump is performed using a WKB approximation. The apparition of a second unstable mode for the streaming instability and its resilience against viscous damping are analysed in Sect.~\ref{sec:results}. Properties of this mode are brought back into the context of planet formation in Sect.~\ref{sec:discussion}.

%----------------------------------------------------------------------------------------
\section{Equations of motion}
\label{sec:motion}

%---------------------------------------------
\subsection{Evolution in a global disc}
\label{sec:global}

\subsubsection{Hypothesis}

We consider non self-gravitating, non magnetic, vertically isothermal discs made of perfect gas. The background surface densities and temperatures of the gas are modelled by power-laws of decreasing exponents, $\Sigma \propto r^{-p}$ and $T \propto r^{-q}$. We use $p = 1$ and $q = 0.4$ to be consistent with models that include detailed radiative transfer (e.g. \citealt{Pinte2014}). Under these assumptions, the gas density $\rhog$ and the pressure $P$ in the midplane of the disc scale as $\rho_{\rm g} \propto r^{- \xi}$ and $P \propto r^{- \xi -q}$, where $\xi \equiv p - q/2 + 3/2$. The effective viscosity of the disc is parametrised using an alpha prescription \citep{Shakura1973}. Numerical simulations of visco-turbulent discs exhibit values of $\alpha \sim 10^{-3} - 10^{-2}$ (e.g. \citealt{Meheut2015}). Dust grains are assumed to be compact, spherical, uncharged, of constant size and density. In observed discs (e.g. \citealt{Williams2014}), local gas surface densities are low enough for dust grains to be in the dilute Epstein drag regime \citep{Epstein1924,Baines1965}. The drag stopping time of the particles is denoted $\tstop$. For millimetre-in-size grains, $\tstop$ is of order of the orbital period at $\sim 50\,$AU \citep{Laibe2012}. The ratio between the drag and the orbital times, often called the Stokes number of the flow, is denoted $\tau_{\rm s}$, consistently with the notations of \citet{Youdin2005}. The dust phase is modelled by a continuous viscousless and pressureless fluid \citep{Saffman1962,Garaud2004}. The local dust-to-gas ratio $\epsilon \equiv \rho_{\rm p} / \rho_{\rm g}$ is larger than the typical $1\%$ of the interstellar medium since dust concentrates vertically and radially in the disc. We follow \citet{Youdin2005} and neglect the vertical stratification of the disc. Our study is therefore relevant for grains with typical sizes $\gtrsim 10 \mu$m that have settled close enough to the midplane \citep{Dubrulle1995,Fromang2009}.
  % a completer

\subsubsection{Pressure maximum}
\label{sec:press_max}

\begin{figure}
	\centering{\includegraphics[width=\columnwidth]{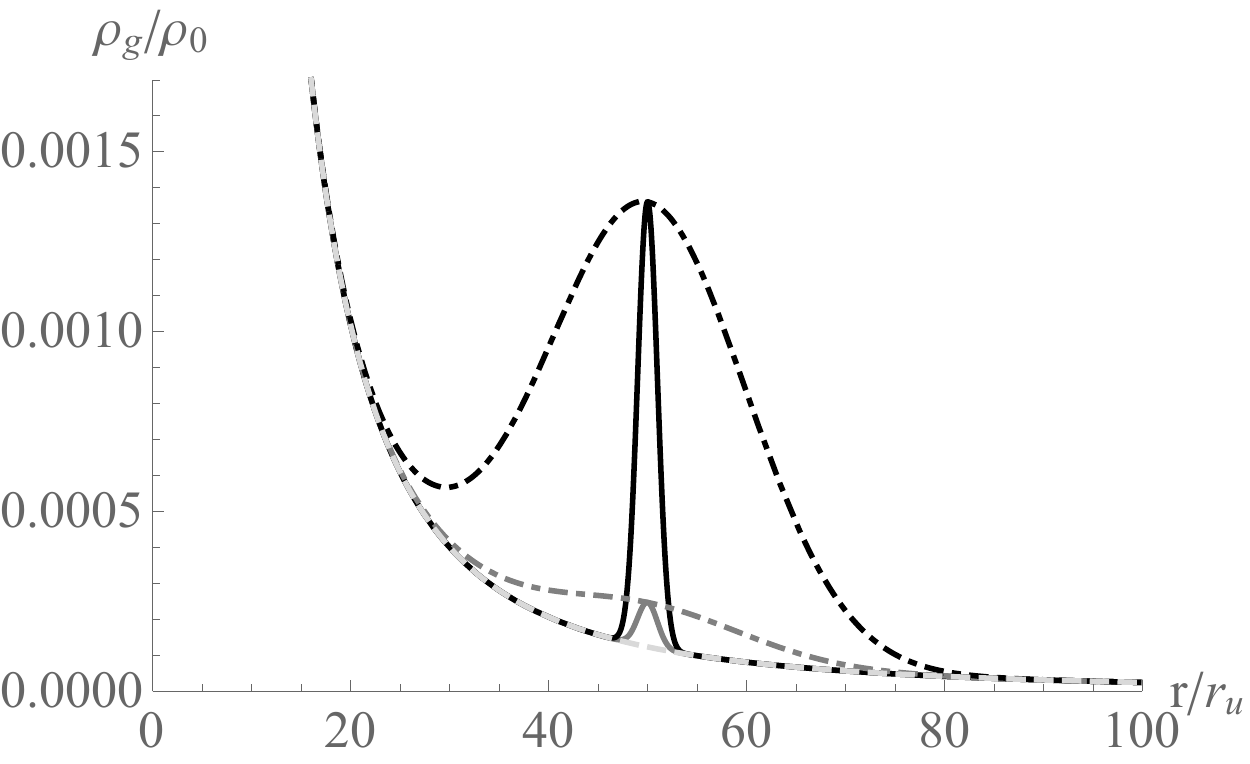}}
	\caption{Zoomed-in density profiles around pressure bumps. Solid and dashed-dotted lines correspond to relative width of the bump $\sigma/r\mrm{u} = 1$ and $\sigma/r\mrm{u} = 10$, while grey and black lines correspond to relative amplitudes $\tilde{A} = 1$ and $\tilde{A} = 10$ respectively. The gas density profile of exponents $p = 1$, $q = 0.4$ in absence of perturbation is given as reference (dashed line/light grey).}
	\label{fig:density}
\end{figure}
We model a local pressure maximum by superimposing a Gaussian perturbation to the usual gas density profile, i.e.
\begin{equation}
	\rhog(r) = \rho_0\left[ \left(\dfrac{r}{r\mrm{u}}\right)^{-\xi} + A \mathrm{e}^{-\dfrac{(r-r_0)^2}{2\sigma^2}} \right]  \, .
	\label{eq:density_max}
\end{equation}
Hence, the amplitude and the width of the density maximum are parametrised by $A$ and $\sigma$ respectively. The radial coordinate is scaled with a radius $r_{\rm u}$ and the Gaussian bump is centred around a position $r_{0}$.  Fig.~\ref{fig:density} shows different shapes of density profiles obtained when varying $A$ and $\sigma$. Note that $A$ should be large enough for the Gaussian perturbation to dominate locally over the decreasing background and thus, for the pressure maximum to exist. The relative amplitude $\tilde{A}$ of the maximum respectively to the background is
\begin{equation}
\tilde{A}\equiv A \left( r_{0}/r_{\rm u}\right)^{\xi} .
\end{equation}
$\tilde{A}$ varies from $\sim 0.1$ for a perturbation due to a Neptune-like mass planet (e.g. \citealt{Dipierro2017}) up to $\sim 10$ for self-induced dust traps \citep{GLM2017}. The width of the bump $\sigma$ is of the order of $\sim H$, the pressure scale height. In absence of any pressure maximum, or when the pressure perturbation is negligible, the orbital correction with respect to a pure Keplerian rotation is of order $\mathcal{O}\left( H^{2}/r^{2} \right)$. In a pressure bump, the orbital correction is of order $\mathcal{O}\left( \tilde{A} \, H^{2}/\sigma^{2}  \right)$ and should remain small enough for the disc to be supported by rotation.

%---------------------------------------------
\subsection{Shearing box approximation}
\label{sec:shearing}
For simplicity, the evolution of gas and dust is studied in a local frame corotating with the disc at a location $\hat{r}_{0}$ and a frequency $\Omega_{0}$. In this shearing box, the coordinates are expanded to the linear order \citep{Goldreich1965}, and $x \equiv r - \hat{r}_{0}$, $y \equiv \hat{r}_{0} \left(\theta - \Omega_{0} t \right)$ and $z$ denote the radial, azimuthal and vertical directions respectively.
\begin{figure}
	\centering{\includegraphics[width=\columnwidth]{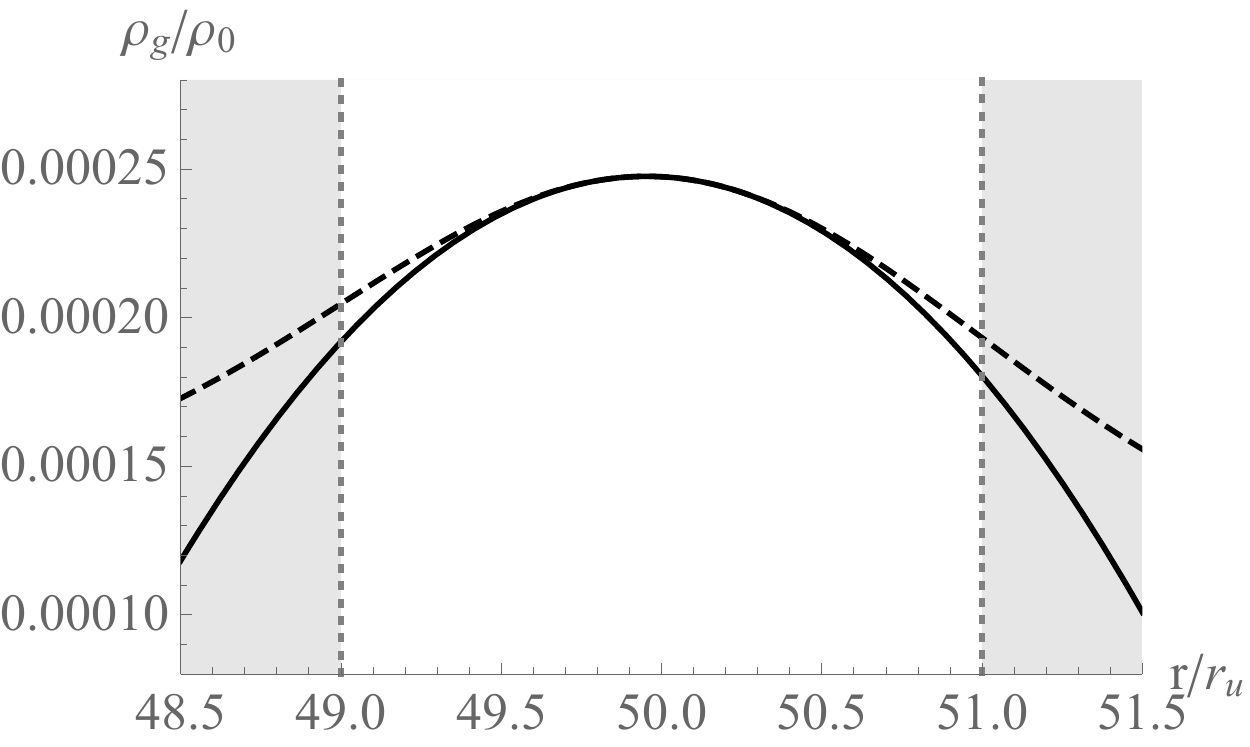}}
	\caption{Comparison between the gas density and its expansion used in the shearing box approximation (dashed and solid lines respectively). Errors are of order $1\%$ in $\mathrm{L}^{1}$ norm, peaking at $5\%$ at the edges of the box of width $\pm\sigma$ delimited by a grey zone. The amplitude and width of the bump are $\tilde{A} = 1$ and $\sigma / r_{\rm u} = 1$.}
	\label{fig:box}
\end{figure}
In the shearing box approximation, the large-scale contribution of the background pressure gradient comes under the form of a constant force  \citep{Youdin2005}. In a box centred around a pressure bump, the large-scale background pressure gradient is a linear function of the distance from the pressure maximum. For convenience, we centre the box around $\hat{r}_{0} = r_{0}$, the maximum of the Gaussian perturbation. Expanding the pressure gradient term to the second order in $x$ provides 
\begin{equation}
	-\dfrac{\bnabla P}{\rhog} \simeq 2r_0\Omega_0^2\left(\eta + \dfrac{\Gamma}{2 r_0}x\right)\vecu_x\,,
	\label{eq:pressure_coeffs}
\end{equation}
with 
\begin{align}
&\eta \equiv     \dfrac{1}{2}  \left. \frac{\mathrm{d} \ln \rhog}{\mathrm{d} \ln r} \right|_{r_{0}}   \left( \dfrac{H_0}{r_0} \right)^2\,, \label{eq:def_eta_gene}\\
&\Gamma \equiv r_0^2  \left. \frac{\mathrm{d}^{2} \ln \rhog}{\mathrm{d} r^{2}} \right|_{r_{0}}  \left(  \dfrac{H_0}{r_0} \right)^2 \label{eq:def_gamma_gene} \,,
\end{align}
where $H_{0}$ denotes the scale height of the disc at $r_{0}$. Following \citet{Youdin2005} notations, we denote $\eta_{0} = \left. \eta \right|_{A = 0}$ the pressure gradient term in absence of bump. Young protoplanetary discs are denser and warmer in the inner regions, which implies $\eta_{0} > 0$. In typical discs, $\eta_{0} \simeq 10^{-2}$. Note that discs with power-law profiles have in general non-zero values for $\Gamma$. This contribution from the curvature of the density profile is neglected in \citet{Youdin2005}, which is an excellent approximation. A pressure bump is defined by $\Gamma > 0$. The pressure maximum position $x_{\rm max} \equiv -2 r_{0}\eta / \Gamma$ is slightly shifted with respect to the centre of the box as a result of the small contribution of the decreasing unperturbed pressure profile. Using Eq.~\ref{eq:density_max} in Eqs.~\ref{eq:def_eta_gene} -- \ref{eq:def_gamma_gene} gives
\begin{align}
	\eta \equiv &  \dfrac{\xi}{2(1+ \tilde{A})}\left( \dfrac{H_{0}}{r_0} \right)^2 , \label{eq:def_eta}\\
	\Gamma \equiv & \left\lbrace - \dfrac{\xi(\xi+1)}{1+\tilde{A}} + \dfrac{\xi^2}{(1+\tilde{A})^2}   \right\rbrace \left( \dfrac{H_{0}}{r_0} \right)^2 + \dfrac{\tilde{A}}{1+\tilde{A}}\left( \dfrac{H_{0}}{\sigma} \right)^2 \, \label{eq:def_gamma}.
\end{align}
The first term of the right-hand side of Eq.~\ref{eq:def_gamma} corresponds to the contribution of the background profile, which increases with $\xi$ (steeper and more curved density profiles). The second term corresponds to the contribution added by the Gaussian perturbation and scales like $\tilde{A}\left( H_{0} / \sigma \right)^{2}$. The relative contribution between the maximum and the background is therefore of order $\tilde{A}\left( r_{0} / \sigma \right)^{2}$.
\begin{figure}
	\centering{\includegraphics[width=\columnwidth]{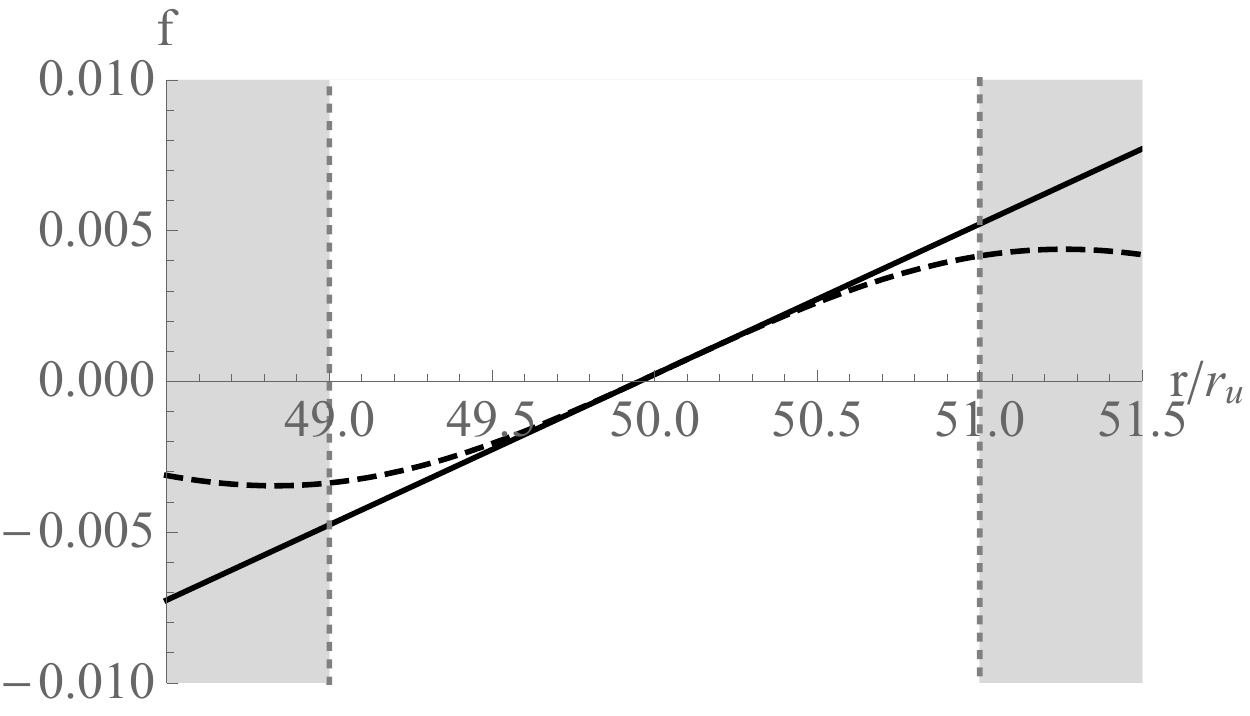}}
	\caption{Similar to Fig.~\ref{fig:box}: comparison between the normalised pressure force (dashed line) and its linearisation expansion (solid line) around $r_{0}$, the maximum of the Gaussian perturbation. Errors are of order $10\%$ in $\mathrm{L}^{1}$ norm, peaking at $20\%$ at the edges of the box.}
	\label{fig:force}
\end{figure}
Fig.~\ref{fig:box} shows that the size of the box is constrained by the validity of the linear approximation. For Eq.~\ref{eq:pressure_coeffs} to remain valid, the size of the box should not exceed $\sim\min \left( \sigma, H\right) $. Note that on the other hand, very narrow bumps may be Rayleigh unstable \citep{Yang2010}. In practice, we use $\sigma  = 1\,r_{\rm u}$. The discrepancy between the real pressure force and its linear approximation is of order $\sim 20 \%$ at most at the edges of the box (Fig.~\ref{fig:force}). Viscous forces are similarly decomposed in large and small scales components. In a typical $\alpha-$disc, the relative contribution between the viscosity and the radial pressure gradient is of order $\alpha \ll 1$ at large scale. This contribution is therefore neglected. On the other hand, the small scale viscous forces, which damp local gas fluctuations, are treated as usual. To avoid unnecessary complications, the local gas sound speed $c_{\rm s}$, the viscosity $\nu = \alpha c_{\rm s} H$, the stopping time of dust grains $\tstop$ and the background dust-to-gas ratio $\epsilon$ are assumed to be constant over the size of the box. This implies that there is more dust in the centre of the bump than at the edges. The equations of motion for the gas and the dust are therefore
\begin{align}
        &\dt{\rhog} + \bnabla\cdot(\rhog\V\mrm{g}) = 0\,, \label{eq:sb1}\\
	&\dt{\rhop} + \bnabla\cdot(\rhop\V\mrm{p}) = 0\,, \label{eq:sb2}
\end{align}	
\begin{align}	
	 \left( \dt{} + \V\mrm{g} \cdot \bnabla \right) \V\mrm{g}  = & -r_0 x \left.\dfrac{\mathrm{d} \OK^2}{\mathrm{d} r}\right|_{r_0} \vecu_x - 2\Omega_0\vecu_z\times\V\mrm{g} \nonumber \\
		&+2r_0\Omega_0^2\left(\eta + \dfrac{\Gamma}{2 r_0}x\right)\vecu_x + \nu\Delta\V\mrm{g}\, \nonumber\\
	& + \dfrac{\rhop}{\rhog}\dfrac{\V\mrm{p}-\V\mrm{g}}{\tstop}\,, \label{eq:sb3} \\
	 \left( \dt{} + \V\mrm{p} \cdot \bnabla \right) \V\mrm{p}  = & -r_0 x \left.\dfrac{\mathrm{d} \OK^2}{\mathrm{d} r}\right|_{r_0} \vecu_x - 2\Omega_0\vecu_z\times\V\mrm{p} \nonumber \\
	&- \dfrac{\V\mrm{p}-\V\mrm{g}}{\tstop}\,. \label{eq:sb4}
\end{align}
In particular, gas and dust have different advection velocities $\V\mrm{g,p}$ and the drag from the dust onto the gas is not neglected \citep{Youdin2005}. For simplicity, physical quantities are used in a dimensionless form, i.e. $\tilde{\omega} \equiv \omega/\Omega_0$, $\tilde{x} \equiv x/\eta_{0}r_0$, $\tilde{k} \equiv k \eta_{0}r_0 $, $\tilde{v} \equiv v/ \eta_{0}r_0\Omega_0$. As a remark, Eqs.~\ref{eq:sb1} -- \ref{eq:sb4} differ from the system studied by \citet{Taki2016}, where the large scale pressure gradient is similar to the constant background introduced in \citet{Youdin2005}, and on top of which a small local Gaussian perturbation of width half of the box is superimposed. This situation corresponds to a pressure perturbation developing at small scales that is weak enough for not affecting background radial velocities of the gas and the dust.

%---------------------------------------------
\subsection{Steady state}
\label{sec:steady}
\begin{figure*}
	\centering{\includegraphics[width=\textwidth]{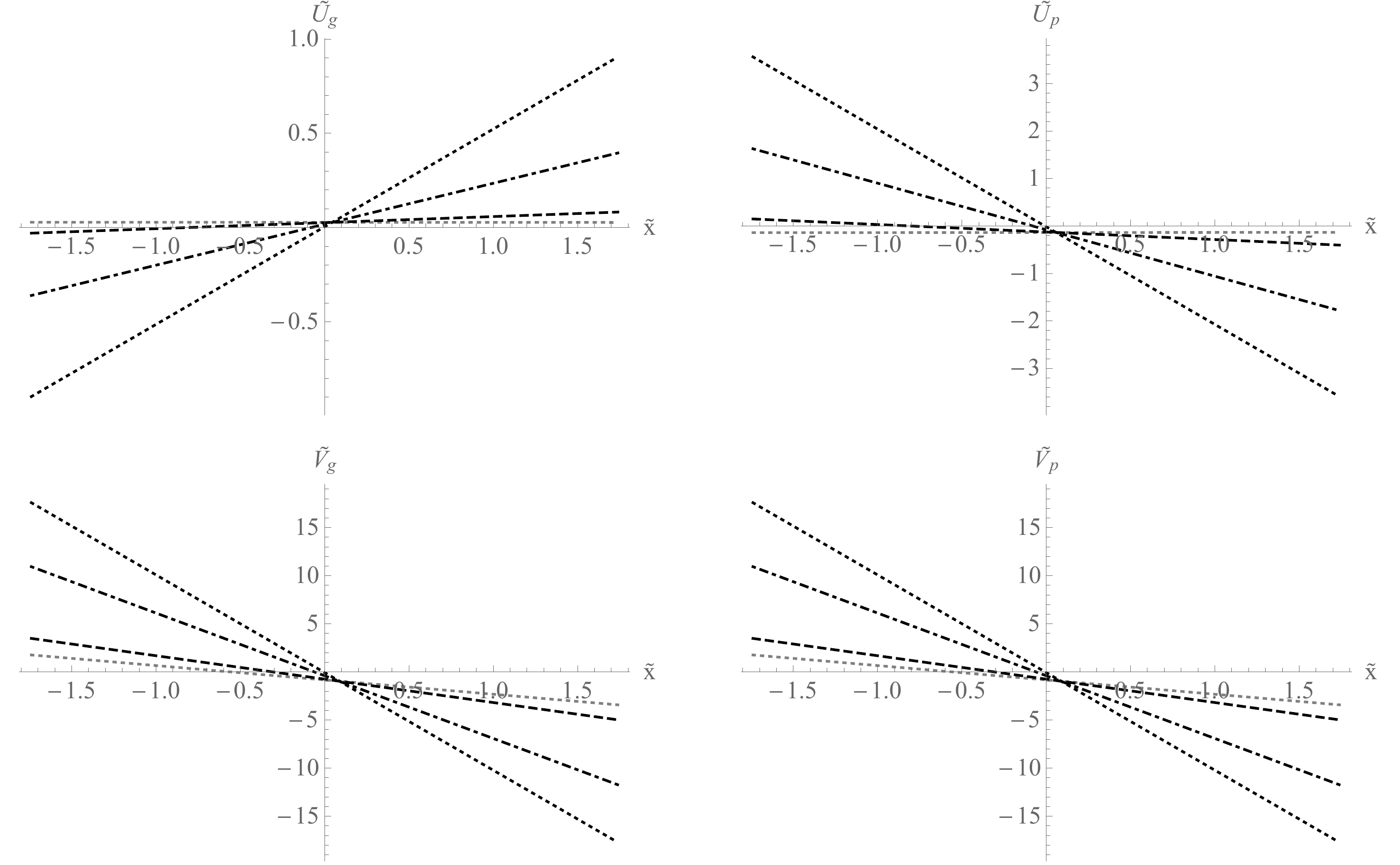}}
	\caption{Radial and azimuthal (top/bottom) velocities for the gas and the dust (left/right) in the box. Bumps with relative amplitudes $\tilde{A} = 0.1 -1 - 10$ are considered (dashed, dot-dashed and dotted black lines), other parameters being those of  the \texttt{linB} problem. Velocities in absence of pressure bump are indicated for reference (grey dotted lines).}
	\label{fig:background}
\end{figure*}
The steady state velocities for gas and dust in the pressure maximum are determined by seeking for solutions of the form $\left( \ol{U}\mrm{g}  , \ol{V}\mrm{g}  \right) = \left( a\mrm{g} x + b\mrm{g} , \alpha\mrm{g} x + \beta\mrm{g} \right)$ and $\left( \ol{U}\mrm{p}  , \ol{V}\mrm{p}  \right) = \left( a\mrm{p} x + b\mrm{p} , \alpha\mrm{p} x + \beta\mrm{p} \right)$. The linear dependancy of the velocities with respect to $x$ is enforced to ensure consistency between the presence of the pressure maximum and the shearing box formalism. Using this Ansatz in Eqs.~\ref{eq:sb1} -- \ref{eq:sb4} provides the \textit{non-linear} system of equations 
\begin{align}
&a\mrm{p}^2 - 3\Omega_0^2 - 2\Omega_0\alpha\mrm{p} + \dfrac{1}{\tstop}(a\mrm{p}-a\mrm{g}) = 0\,, \label{eq:1} \\
&a\mrm{p}b\mrm{p} - 2\Omega_0\beta\mrm{p} + \dfrac{1}{\tstop}(b\mrm{p} - b\mrm{g}) = 0\,, \label{eq:2}\\
&a\mrm{p}\alpha\mrm{p} + 2\Omega_0 a\mrm{p} + \dfrac{1}{\tstop}(\alpha\mrm{p}-\alpha\mrm{g}) = 0\,, \label{eq:3} \\
&b\mrm{p}\alpha\mrm{p} + 2\Omega_0 b\mrm{p} + \dfrac{1}{\tstop}(\beta\mrm{p} - \beta\mrm{g}) = 0\,, \label{eq:4} \\
&a\mrm{g}^2 - 3\Omega_0^2 - 2\Omega_0\alpha\mrm{g} - \dfrac{\epsilon}{\tstop}(a\mrm{p} - a\mrm{g}) - \Gamma\Omega_0^2 = 0\,, \label{eq:5} \\
&a\mrm{g}b\mrm{g} - 2\Omega_0\beta\mrm{g} - \dfrac{\epsilon}{\tstop}(b\mrm{p}-b\mrm{g}) - 2r_0\Omega_0^2\eta = 0\,, \label{eq:6} \\
&a\mrm{g}\alpha\mrm{g} + 2\Omega_0 a\mrm{g} - \dfrac{\epsilon}{\tstop}(\alpha\mrm{p} - \alpha\mrm{g}) = 0\,, \label{eq:7} \\
&b\mrm{g}\alpha\mrm{g} + 2\Omega_0 b\mrm{g} - \dfrac{\epsilon}{\tstop}(\beta\mrm{p} - \beta\mrm{g}) = 0\,. \label{eq:8}
\end{align}
These non-linear terms originate from non-trivial advection terms specific to the pressure bump and require care for numerical root finding (see Appendix~\ref{sec:vel} for technical details). Neglecting this additional advection provides a crude approximation of the solution of Eqs.~\ref{eq:1} -- \ref{eq:8} by taking the solution of \citet{NSH1986} given in Appendix~\ref{sec:vel} and replacing $\eta$ by $\eta + \frac{\Gamma}{2 r_{0}}x$. We find errors of order $\sim10\%$ up to $\sim100\%$ between the two approaches, a discrepancy becoming important for the radial velocity of the gas. Fig.~\ref{fig:background} illustrates the dust and the gas motion inside the bump for various relative amplitudes of the pressure maximum. As a reminder, pure Keplerian shear is $v_{y} = -\frac{3}{2} \Omega_{0} x$. In absence of a bump, dust (resp. gas) drifts inwards (resp. outwards) by conservation of angular momentum. Both the gas and the dust are sub-Keplerian. Inside a bump, dust drifts towards the pressure maximum, while gas drifts outwards. This requires for gas and dust to orbit at super-Keplerian frequency in the inner edge of the bump. Both the gas and the dust radial velocities are rigorously zero at $x_{\rm max} < 0$ the location of the pressure maximum. Although the different velocities appear to cross each other at the same location in Fig.~\ref{fig:background}, this is actually not the case. Mathematically, the intersection between the line corresponding to $\tilde{A} = 0$ and the other lines depends slightly on the different physical parameters. In particular, for increasing values of $\Gamma$, the intersecting point becomes closer to the centre of the box.

%---------------------------------------------
\subsection{Perturbation}
\label{sec:perturb}

The linear stability of the system Eqs.~\ref{eq:sb1} -- \ref{eq:sb4} is investigated by looking for perturbations of the form
\begin{align}
\label{perturb}
&\rhog = \rhog^0(1+\delg)\,, \\
&\rhop = \rhop^0(1+\delp)\,, \\
&\V\mrm{g} = \ol{\V}\mrm{g} + \vecuv\mrm{g} = \ol{\V}\mrm{g} + u\mrm{g}\vecu_x + v\mrm{g}\vecu_y + w\mrm{g}\vecu_z \,, \\
&\V\mrm{p} = \ol{\V}\mrm{p} + \vecuv\mrm{p} = \ol{\V}\mrm{p} + u\mrm{p}\vecu_x + v\mrm{p}\vecu_y + w\mrm{p}\vecu_z \,,
\end{align}
where
\begin{equation}
	\delta(x,z,t) = \Delta(x)e^{i(k_x x + k_z z - \omega t)}\,.
	\label{eq:delta}
\end{equation}
In absence of a pressure bump, the perturbation must develop in both the $x$ and the $z$ direction to become unstable \citep{Youdin2005,Jacquet2011}. This property originates from local conservation of the gas mass. With a pressure bump, advection terms enforce the amplitude of the perturbation $\Delta$ to depend on $x$. For simplicity, we focus on cases where this amplitude varies slowly compared to the phase ($Hk_{x}\gg 1$) and use a WKB approximation to compute spatial derivatives, i.e.
\begin{equation}
\frac{\partial \delta(x,z,t)}{\partial x}  \simeq  i k_{x} \Delta(x) e^{i(k_x x + k_z z - \omega t)} .
\label{eq:anzatz_pert}
\end{equation}
We obtain the following set of 8 linear equations for the perturbation
\begin{align}
	-i\omega\delg &+ i\delg k_x\ol{U}\mrm{g} + \delg a\mrm{g} + ik_x u\mrm{g} + ik_z w\mrm{g} = 0\,, \label{eq:pert_rhog} \\
	-i\omega\delp &+ i\delp k_x\ol{U}\mrm{p} + \delp a\mrm{p} + ik_x u\mrm{p} + ik_z w\mrm{p} = 0\,, \label{eq:pert_rhop} \\
	\nonumber -i\omega\vecuv\mrm{g} &+ i k_x\ol{U}\mrm{g}\vecuv\mrm{g} + (u\mrm{g} a\mrm{g} - 2v\mrm{g}\Omega_0)\vecu_x + u\mrm{g}(\alpha\mrm{g} + 2\Omega_0)\vecu_y \\
	&-\dfrac{\epsilon}{\tstop}(\vecuv\mrm{p} - \vecuv\mrm{g} + (\delp-\delg)(\ol{\V}\mrm{p} - \ol{\V}\mrm{g})) \nonumber \\
	&+ i c_s^2\delg\veck +\nu(k_x^2 + k_z^2)\vecuv\mrm{g} = 0\,, \label{eq:pert_vg} \\
	\nonumber -i\omega\vecuv\mrm{p} &+ i k_x\ol{U}\mrm{p}\vecuv\mrm{p} + (u\mrm{p}a\mrm{p} - 2v\mrm{p}\Omega_0)\vecu_x + u\mrm{p}(\alpha\mrm{p} + 2\Omega_0)\vecu_y \nonumber \\
	&+\dfrac{1}{\tstop}(\vecuv\mrm{p} - \vecuv\mrm{g}) = 0\, . \label{eq:pert_vp} 
\end{align}
where the four background velocities $\ol{U}\mrm{g}, \ol{U}\mrm{p}, \ol{V}\mrm{g}, \ol{V}\mrm{p}$ are the linear functions determined in Sect.~\ref{sec:steady} and not the solutions derived by \citet{NSH1986}. In practice, the eigenmodes $\omega_{1,8}$ of the system Eqs.~\ref{eq:pert_rhog} -- \ref{eq:pert_vp} are determined by finding zeros of the determinant of the perturbation matrix numerically, using a sufficient precision. Note that the eigenmodes obtained by this procedure depend on $x$, i.e. $\omega_{1,8} (x)$, which may sound inconsistent with Eq.~\ref{eq:delta}. However, at small times,
\begin{equation}
k_{x}^{-1} \frac{\mathrm{d} \omega}{\mathrm{d}x} t \simeq \frac{\Omega_{0} t}{k_{x} H}.
\end{equation}
In this case, the solution is consistent with the initial Anzatz and the WKB approximation over a number $n = k_{x} H \gg 1$ of orbital periods. To compare our results with the case of an inviscid disc with no pressure bump, we use the test cases \texttt{linA, linB, linC, linD} studied in \citet{Youdin2007,Bai2010} (see parameters in Appendix~\ref{sec:lin}). Our procedure provides the expected coefficients with similar precision.

%----------------------------------------------------------------------------------------
\section{Results}
\label{sec:results}

\subsection{Unstable modes}
\label{sec:unstable}

\begin{figure}
	\centering{\includegraphics[width=\columnwidth]{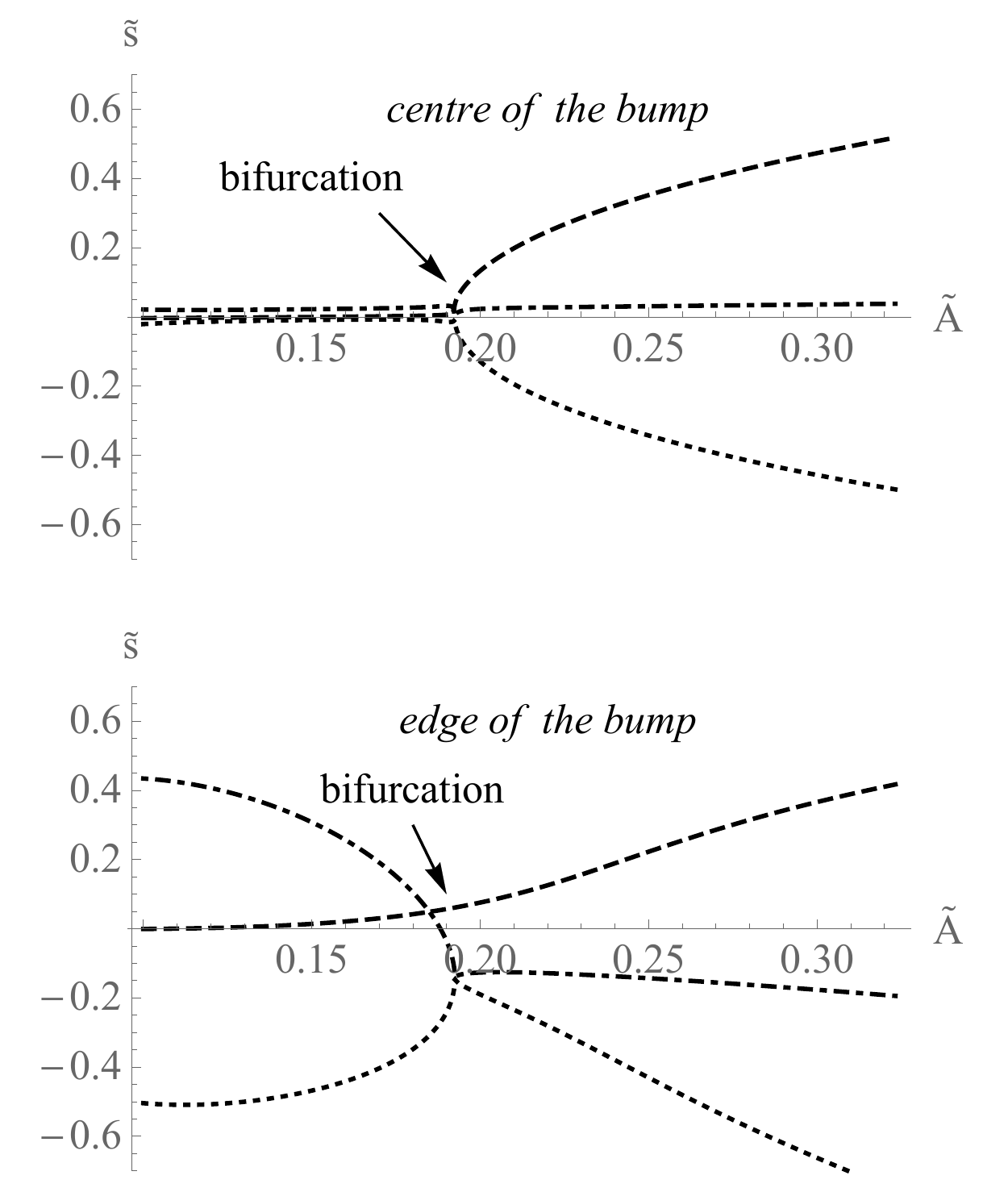}}
	\caption{Growth rates of the two epicyclic modes and the secular mode of the streaming instability for increasing values of the relative amplitude $\tilde{A}$ of the bump. In absence of any pressure maximum, the only unstable mode is the usual secular mode (dot-dashed/black). In pressure maxima, a novel unstable mode develops (dashed/black). A bifurcation between the secular mode and the other epicycle (dotted/black) occurs at $\tilde{A}\simeq 0.2$. Parameters correspond to the \texttt{linA} problem.}
	\label{fig:modes_A}
\end{figure} 

%%%%%%%%%%%%%%%%%%%%%%%% ICI %%%%%%%%%%%%%%%%%%%%%%%%%

The growth rates $s$ obtained from the procedure described in Sect.~\ref{sec:perturb} are shown in Fig.~\ref{fig:modes_A}. Three modes are considered: two modified epicyclic modes and the secular mode of the streaming instability identified in \citet{Youdin2005}, the other modes playing no particular role in this problem. For tiny perturbations of the pressure profile, the only unstable mode is the secular mode, similarly to what happens in a disc with no bump. Novel features appear when increasing progressively the relative amplitude of the bump. Fig.~\ref{fig:modes_A} shows that one of the two epicyclic modes becomes unstable. The related growth rate may become larger than the growth rate of the secular mode as the amplitude of the bump increases. The possible instability of the epicyclic modes was mentioned in \citet{Youdin2005}. When increasing further the amplitude of the bump, a transition between two distinct regimes occurs for relative amplitudes of order $\tilde{A} \gtrsim 0.2$, a conservative value for typical bumps. Fig.~\ref{fig:modes_A} shows that a bifurcation between the secular and the stable epicyclic modes that gives birth to two new modes. We find that this bifurcation is universal and does in particular not depend on the dust-to-gas ratio. The exact number of unstable modes depends on the distance to the pressure maximum. Near $x_{\rm max}$, two modes are unstable, whereas at the edges of the bump, only one mode is unstable. This mode corresponds to the novel unstable mode originating from the bifurcation and is not the secular mode of the streaming instability which grows in absence of maximum. Physically, this novel instability originates from the strong differential advection between the gas and the dust at the edges of the bump powered by the background velocities in the bump.

\begin{figure}
	\centering{\includegraphics[width=\columnwidth]{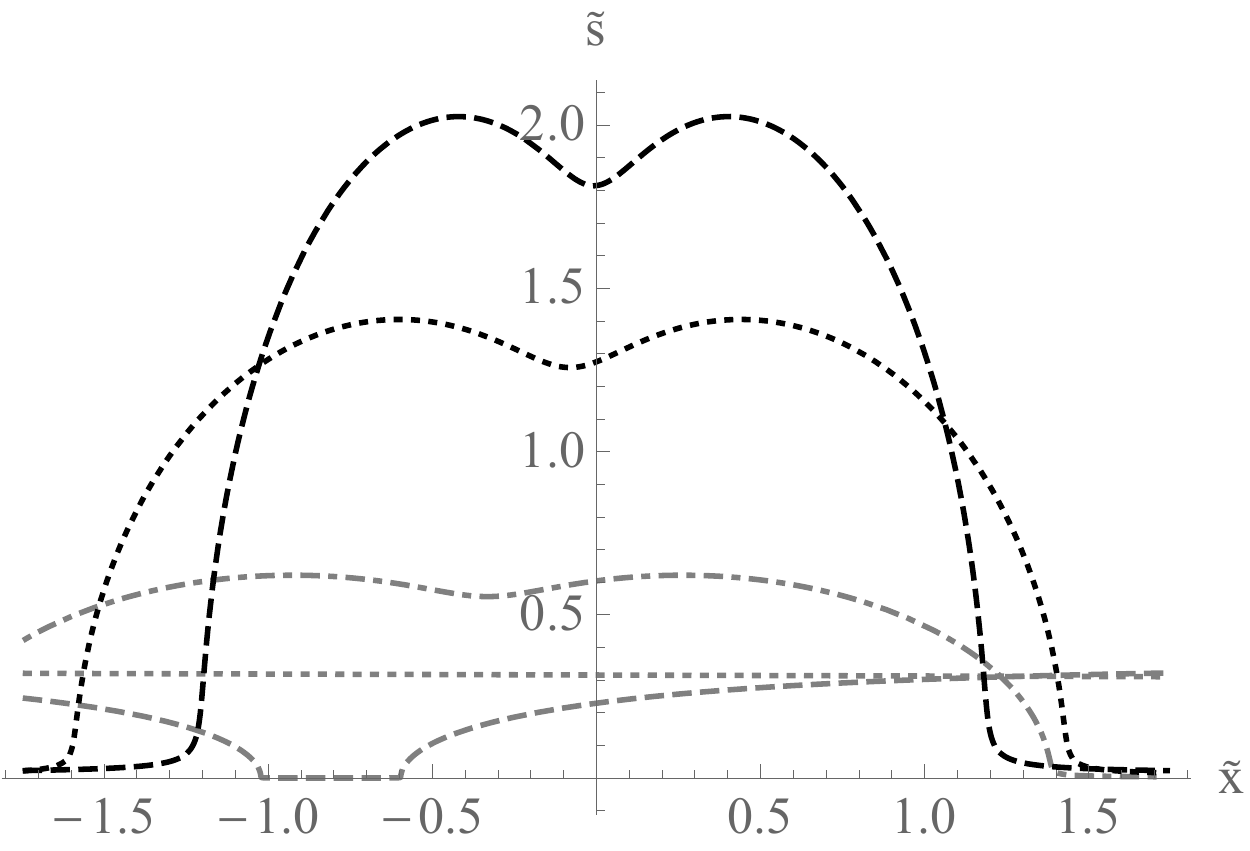}}
	\caption{Growth rate of the streaming instability in the box for bumps of relative amplitudes $\tilde{A} = 0.1 - 0.25 - 1 - 10 $ (dashed/grey, dot-dashed/grey, dotted/black and dashed/black respectively). For shallow bumps, streaming instability develops everywhere except near the maximum where the pressure gradient is zero. For large bumps, the instability develops more efficiently close to the maximum. Parameters are those of the $\texttt{linD}$ problem. The growth rate in absence of maximum is given by the dotted/grey line.}
	\label{fig:growth_D}
\end{figure}

Fig.~\ref{fig:growth_D} shows how the growth rates depend on the radial location in the bump for increasing amplitudes, showing the consequences of the bifurcation identified above. Rigorously, the exact location of the pressure maximum depends on $\tilde{A}$ and is slightly offset from the centre of the shearing box due to the local curvature of the pressure profile in absence of maximum (see Sect.~\ref{sec:steady}). For shallow bumps ($\tilde{A} \lesssim 0.2$), streaming instability develops everywhere except close to the pressure maximum. Its efficiency is maximum at the edges of the bump, where the local pressure gradient is the greatest. This behaviour is consistent with the linear analysis of \citet{Youdin2005}, valid for discs with monotonic pressure profiles. For larger amplitudes ($\tilde{A} \gtrsim 0.2$), the streaming instability grows more efficiently in the centre of the bump than at the edges. As shown in Fig.~\ref{fig:growth_D}, its efficiency is slightly reduced at $x = x_{\rm max}$. At the edges of the bump, the growth rate decreases up to eventually reach zero. No instability develops in this particular case. Hence, for a shallow bump in an inviscid disc, gas and dust are linearly unstable everywhere except at the exact location of the pressure maximum. On the opposite, when the disc contains a bump that is large enough (i.e. for pressure perturbation to be of a least a few ten percents), dust concentration may occur preferentially inside the maximum, and not at the edges, as one would have expected with the classical linear stability analysis.
\begin{figure}
	\centering{\includegraphics[width=\columnwidth]{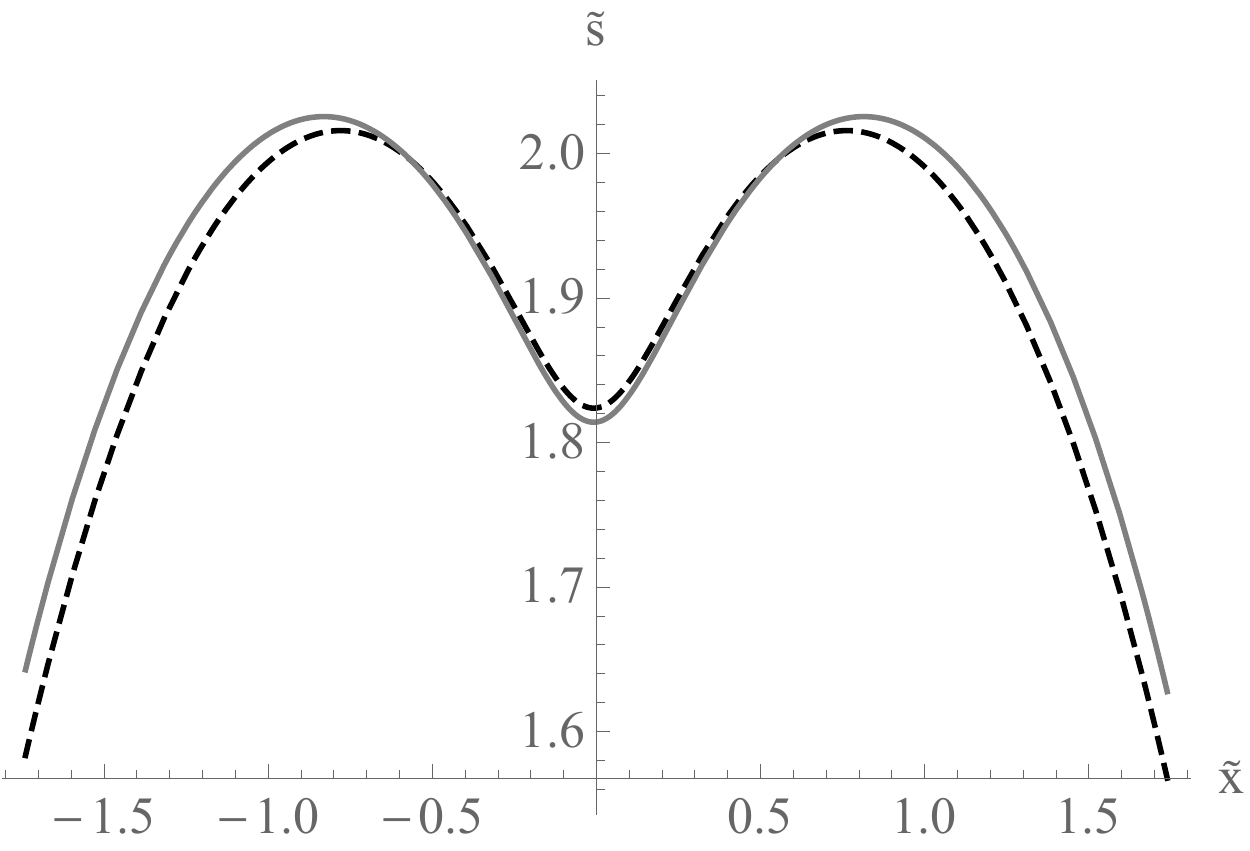}}
	\caption{Growth rate of the instability for two same ratios $\lambda/l_{\rm stop} = 1$. Growth rates are almost similar, discrepancies of a few percent originating from slightly different advection velocities. We use $\tilde{A}=10$, $\epsilon=2$, $\tau\mrm{s}=0.01$, $\tilde{k}_{x,z} = 100$ (black dashed line) and $\tau\mrm{s}=0.001$, $\tilde{k}_{x,z} = 1000$ (grey solid line).}
	\label{fig:equiv}
\end{figure}

\begin{figure}
	\centering{\includegraphics[width=\columnwidth]{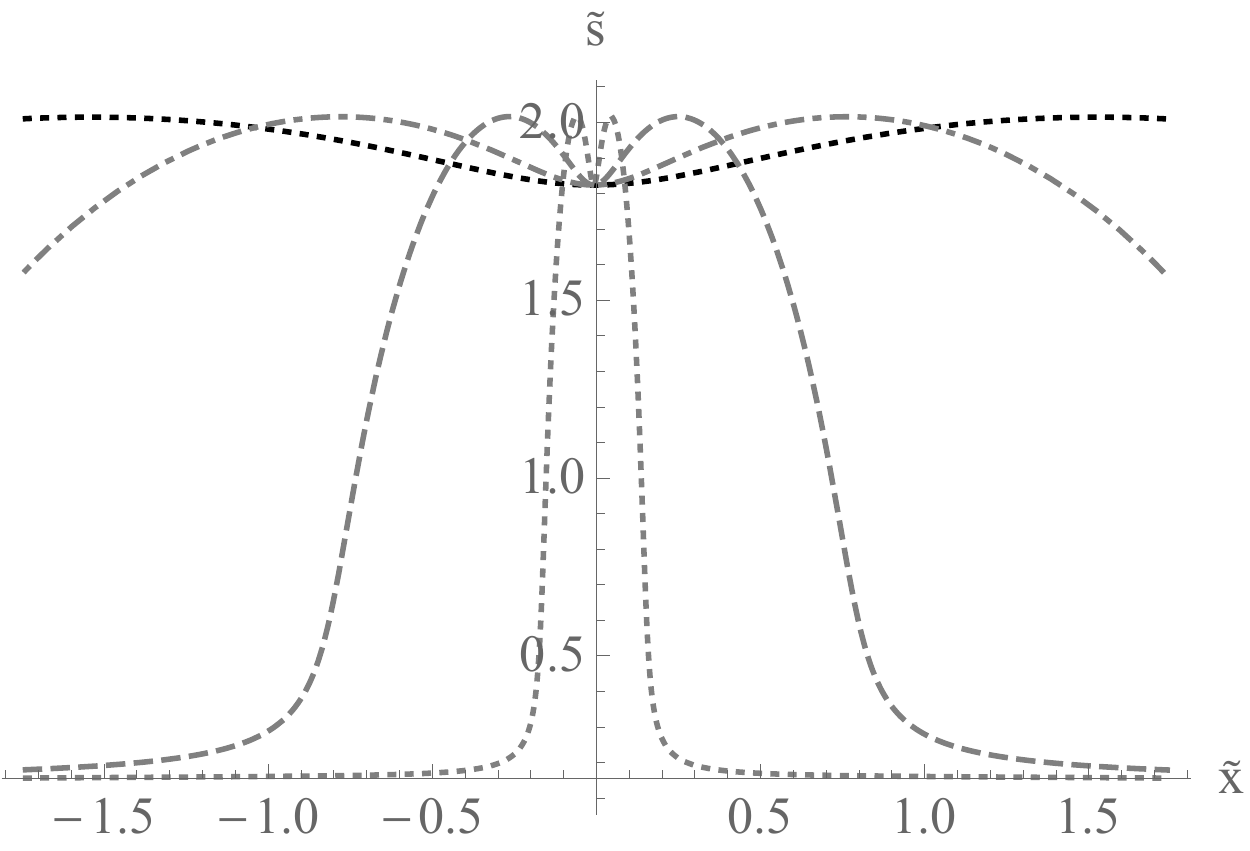}}
	\caption{Growth rates for a bump of relative amplitude $\tilde{A}=10$ and different wavenumbers $\tilde{k}_{x,z}=50-100-300-1500$ (dotted/black, dot-dashed/grey, dashed/grey, dotted/grey). The region where the instability develops is more extended for larger wavelengths. Parameters are those of the \texttt{linC} problem.}
	\label{fig:wavelength}
\end{figure}

The growth rates depend on the ratio between the wavelength $\lambda$ of the perturbation and $l_{\rm stop}~\equiv~\eta_{0} r_{0} \Omega_{0} \tstop$, the length over which the gas decouples from the dust, sometimes referred as the stopping length. The ratio $\lambda / l_{\rm stop}$ measures the number of perturbations over which the stopping length spreads. Fig.~\ref{fig:equiv} shows that for similar values of $\lambda / l_{\rm stop}$ or equivalently $\tilde{k}_{x,z} \, \tau_{\rm s}$, the growth rates obtained are almost identical. Corrections of order a few percents are due to slightly different values for the advection velocities. Fig.~\ref{fig:wavelength} shows growth rates obtained for different wavelengths $\lambda$. For $l_{\rm stop} \gg \lambda $, dust and gas experience the details of the pressure profile before being coupled together by the drag, and the growth rate profile is narrow. Instead, for $l_{\rm stop} < \lambda $, dust and gas are quickly coupled by the drag, their differential velocity is proportional to the local pressure gradient, and this information is carried away by the perturbation. In this case the growth rate profile is wider. The ratio $\lambda/l\mrm{stop}$ sets the width of the region where no instability develops for shallow bumps, and the width of the central region where the instability is weakened for large bumps.

\subsection{Viscosity}
\label{sec:viscosity}
\begin{figure}
	\centering{\includegraphics[width=\columnwidth]{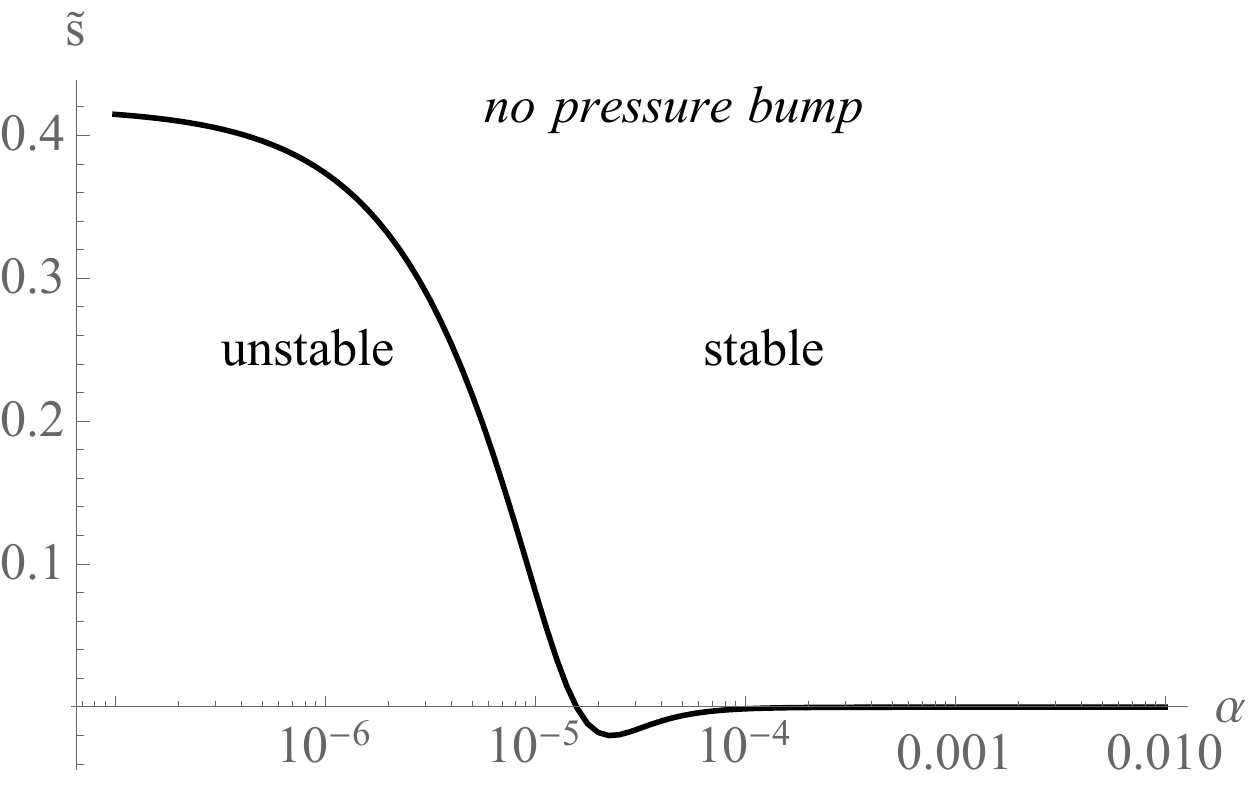}}
	\caption{Growth rate of the instability in absence of bump for increasing viscosities. The unstable mode is damped for $\alpha\gtrsim 10^{-5}$. Parameters correspond to the \texttt{linA} problem.}
	\label{fig:visc_nomax}
\end{figure}
\begin{figure}
	\centering{\includegraphics[width=\columnwidth]{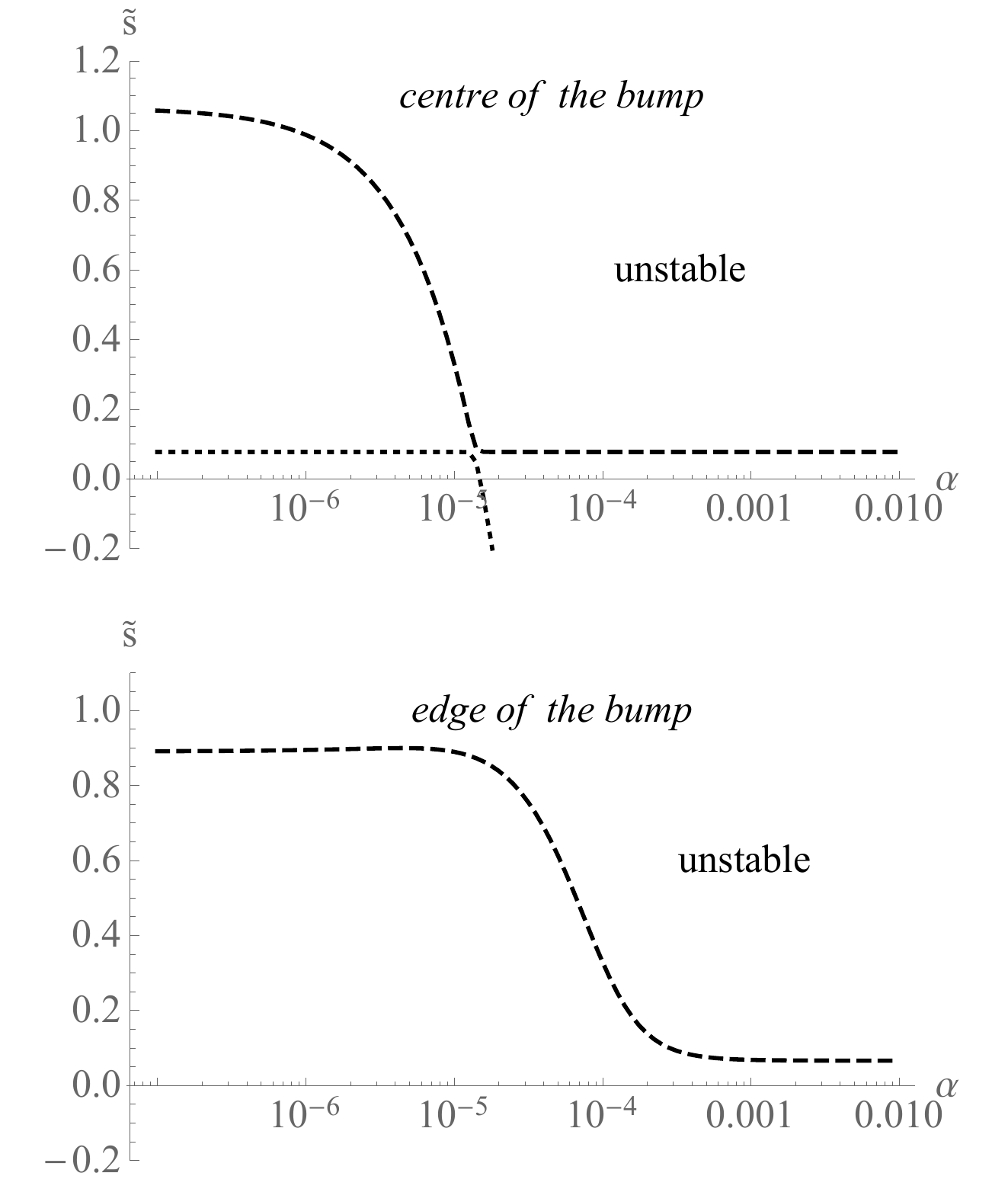}}
	\caption{Growth rates of the unstable modes for increasing viscosities. An unstable mode exists even in highly viscous discs. Its growth rate is weakened by one order of magnitude compared to the inviscid case. Parameters correspond to the \texttt{linA} problem.}
	\label{fig:bifurcation}
\end{figure}
\begin{figure}
	\centering{\includegraphics[width=\columnwidth]{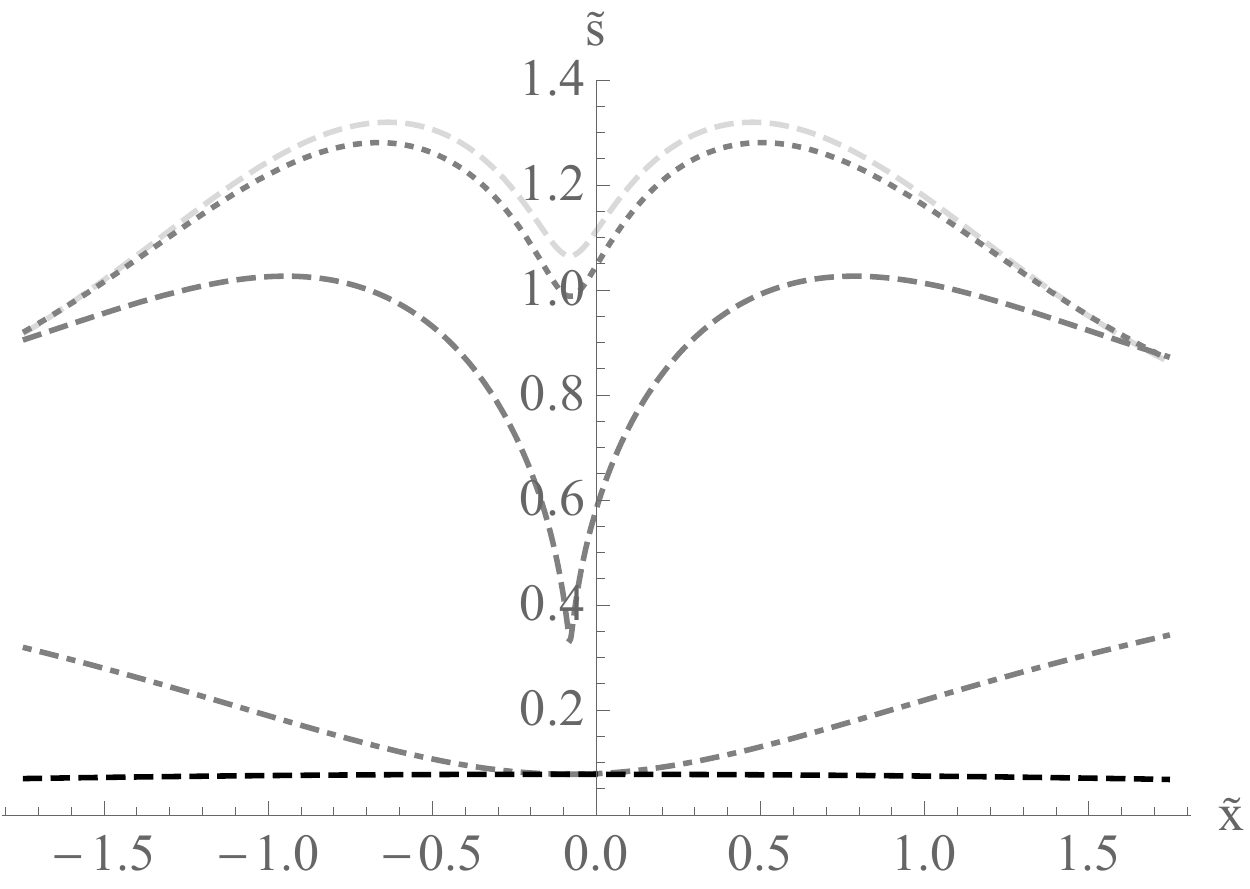}}
	\caption{Growth rates for viscous parameters $\alpha = 0-10^{-6}-10^{-5}-10^{-4}-10^{-3}$ (dashed/light, dotted, dashed, dash-dotted grey and dashed/black lines respectively). Parameters are those of the \texttt{linA} problem, the relative amplitude of the bump is $\tilde{A} = 1$.}
	\label{fig:viscosity}
\end{figure}
\begin{figure}
	\centering{\includegraphics[width=\columnwidth]{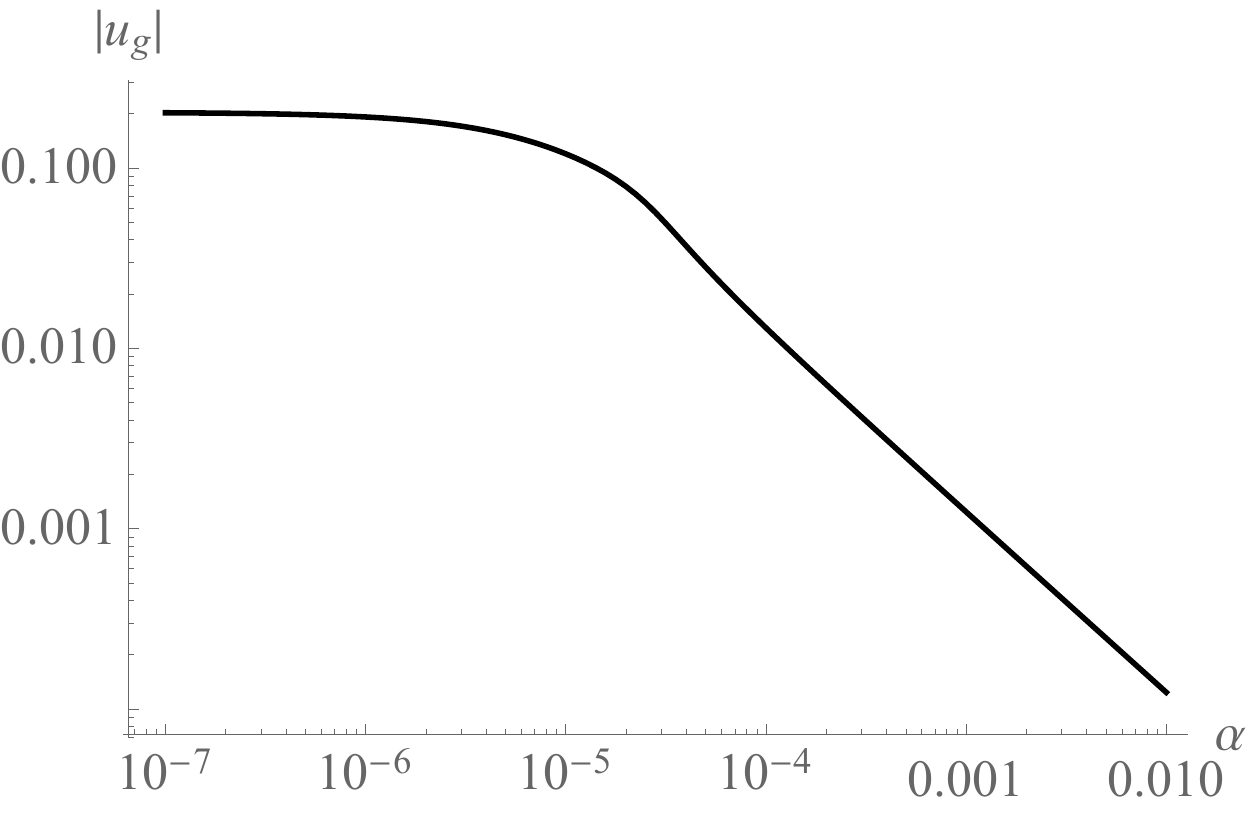}}
	\caption{Amplitude of the perturbation of the radial velocity of the gas $|u_{\mathrm{g}}|$ for increasing viscosities. The fluctuations are damped efficiently at large viscosities. The instability does not require fluctuations in the gas velocity to grow, since differential back-reaction is provided by the bump itself. Parameters are those of the \texttt{linA} problem.}
	\label{fig:norme}
\end{figure}
\begin{figure}
	\centering{\includegraphics[width=\columnwidth]{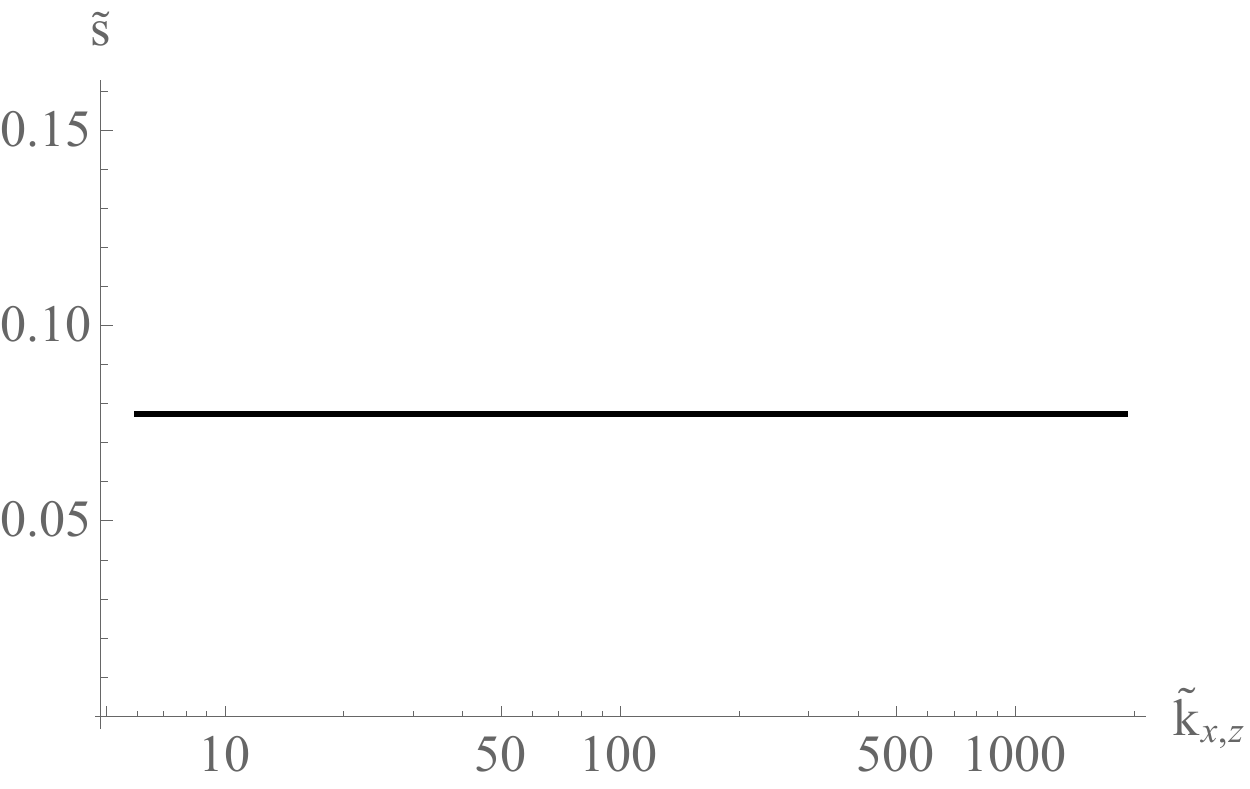}}
	\caption{Growth rate of the instability for different wavenumbers $\tilde{k}_{x,z}$ in a viscous disc with $\alpha=10^{-2}$. No dependancy is found, consistently with a mechanism powered by the background profile of the bump. Parameters are those of the \texttt{linA} problem.}
	\label{fig:depk}
\end{figure}
\begin{figure}
	\centering{\includegraphics[width=\columnwidth]{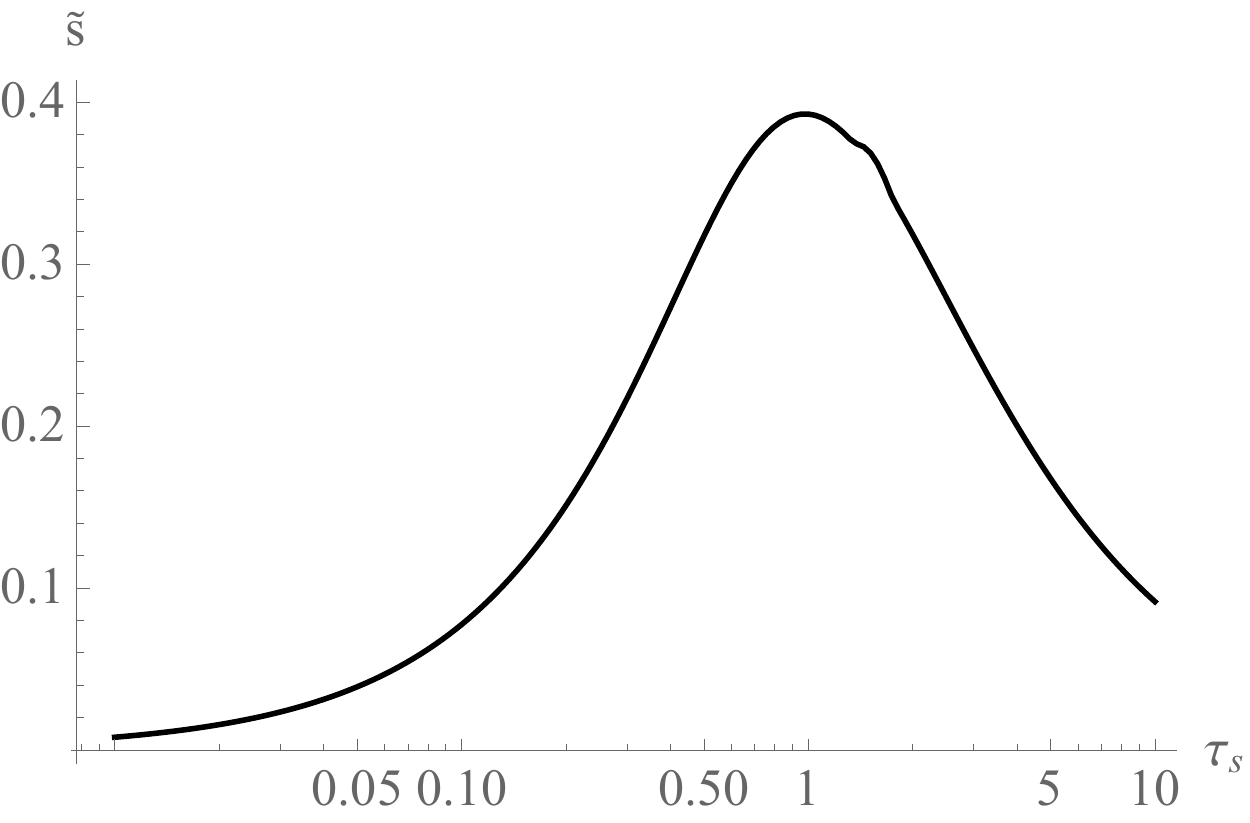}}
	\caption{Growth rate of the instability as a function of the Stokes number $\tau\mrm{s}$ for $\alpha=10^{-2}$. Maximal efficiency is reached for $\tau\mrm{s}\sim 1$ as expected. Parameters are those of the \texttt{linA} problem.}
	\label{fig:taus}
\end{figure}
\begin{figure}
	\centering{\includegraphics[width=\columnwidth]{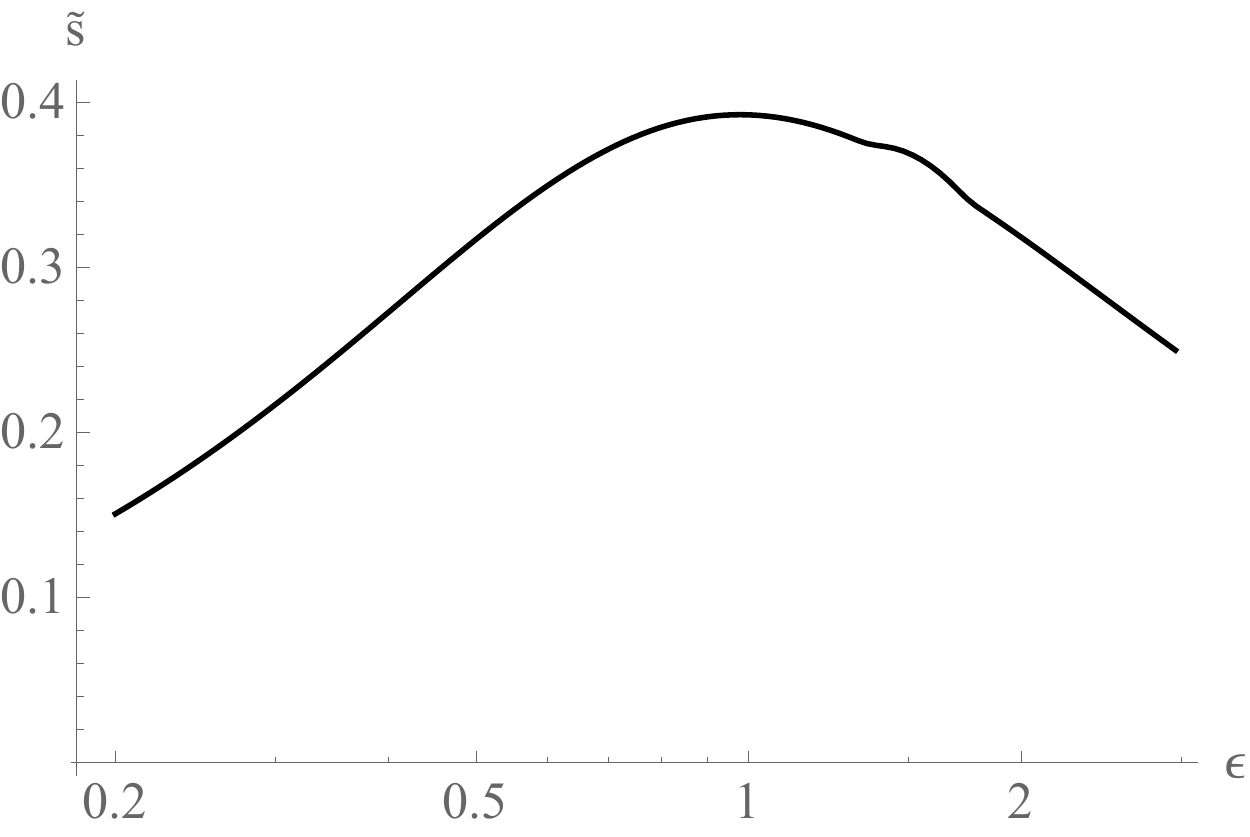}}
	\caption{Growth rate of the instability as a function of the dust-to-gas ratio $\epsilon$ for $\alpha=10^{-2}$. Maximal efficiency is reached for $\epsilon\sim 1$ as expected. Parameters are those of the \texttt{linA} problem.}
	\label{fig:epsilon}
\end{figure}
\begin{figure*}
	\centering{\includegraphics[width=\textwidth]{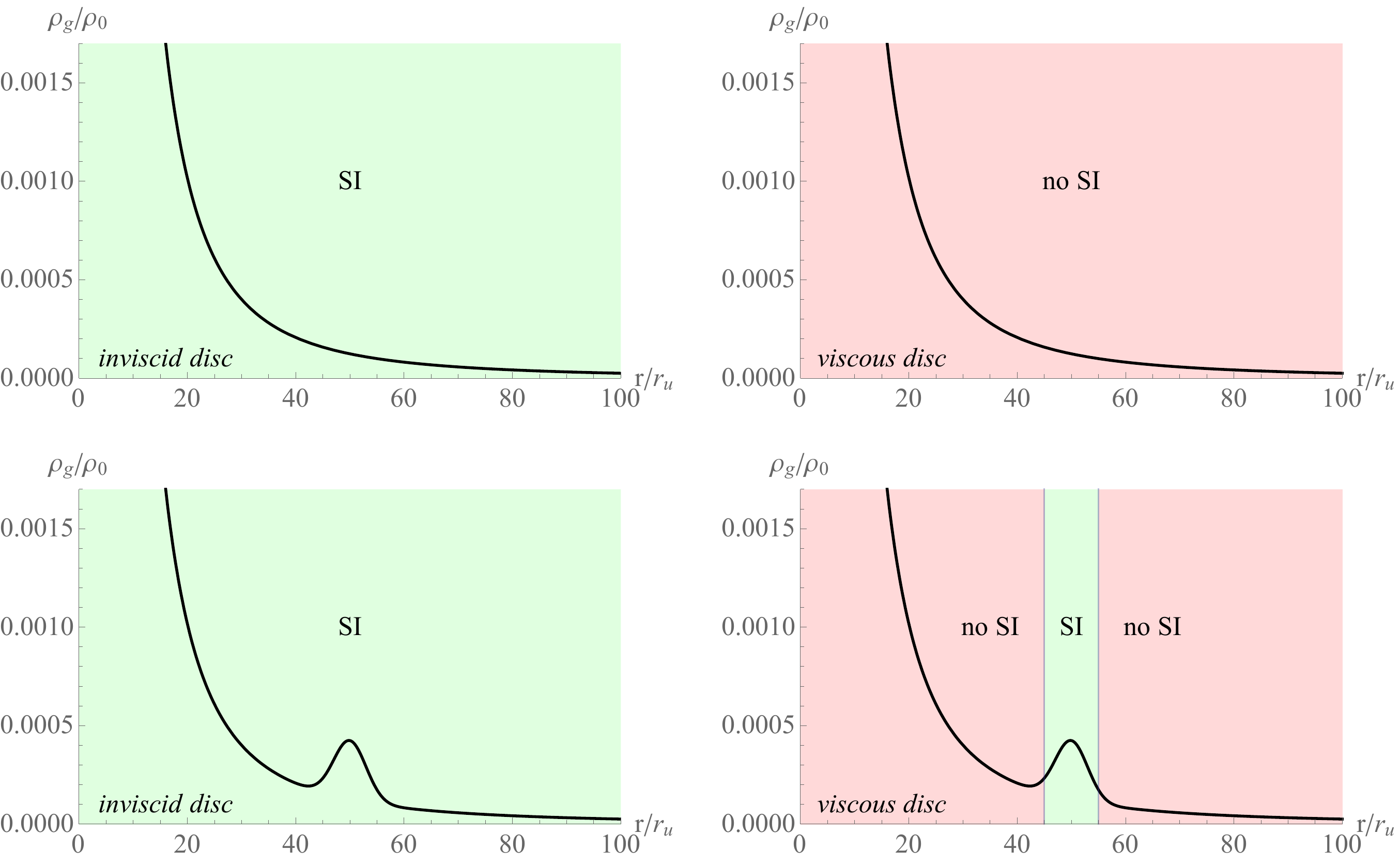}}
	\caption{Sketch illustrating where the streaming instability grows in different discs. An inviscid disc is linearly unstable everywhere, with or without a pressure bump. In visco-turbulent discs, the streaming instability develops only inside pressure bumps.}
	\label{fig:concl_growth}
\end{figure*}

The ability for viscosity to damp the instability is now investigated. Fig.~\ref{fig:visc_nomax} shows that in absence of a pressure bump, viscosity prevents the development of the streaming instability for values of $\alpha$ as low as $ \sim 10^{-5}$. This value depends on the wavelength of the perturbation, smaller fluctuations being damped more efficiently by viscosity. Streaming instability does therefore not grow in typical visco-turbulent discs where $\alpha \sim 10^{-3} - 10^{-2}$. However, when a significant bump is present in the disc, a different behaviour is observed. Fig.~\ref{fig:bifurcation} shows that for increasing values of $\alpha$, one of the two unstable modes is suppressed, whereas the other one is only weakened, \textit{but not damped}. The growth rate is reduced by one order of magnitude compared to the inviscid case (see Fig.~\ref{fig:viscosity}). Still, this mode develops in a time relevant for planetesimal formation (see Sect.\ref{sec:discussion}). Hence, streaming instability is found to always develop in a pressure bump, even in highly viscous discs. The fact that viscosity does not entirely damp the instability in the bump may appear counterintuitive. This apparent conundrum can be explained by noting that viscosity damps only the perturbations of the gas velocity. Fig.~\ref{fig:norme} shows that these perturbations are actually suppressed more and more efficiently when viscosity increases. However, in pressure bumps, the required gradients of back-reaction are provided by the background velocities resulting from the local pressure profile. Hence, the instability can grow even if perturbations in the gas velocity are damped by viscosity. This effect is not observed in a disc with monotonic pressure profiles, since local gradients of back-reaction originate only from perturbations of the gas velocity that are killed by viscosity. Consistently, growth rates in the viscous regime do almost not depend on the wavelength, as shown in Fig.~\ref{fig:depk}. Fig.~\ref{fig:taus} and Fig.~\ref{fig:epsilon} show that similarly to the usual case, streaming instability is most efficient for $\tau\mrm{s} \sim 1$ and $\epsilon \sim 1$, such parameters being typical protoplanetary discs.
Fig.~\ref{fig:concl_growth} summarises where streaming instability develops in discs, whether a pressure bump is present or not and whether the disc is inviscid or viscous. A dusty disc with monotonic pressure profile is linearly unstable everywhere if it is inviscid, and stable everywhere if it is viscous \citep{Youdin2005}. The latter is not true anymore in presence of a pressure maximum: streaming instability always grows in the bump, even if the disc is viscous.
%

%----------------------------------------------------------------------------------------
\section{Discussion}
\label{sec:discussion}
\begin{figure}
	\centering{\includegraphics[width=\columnwidth]{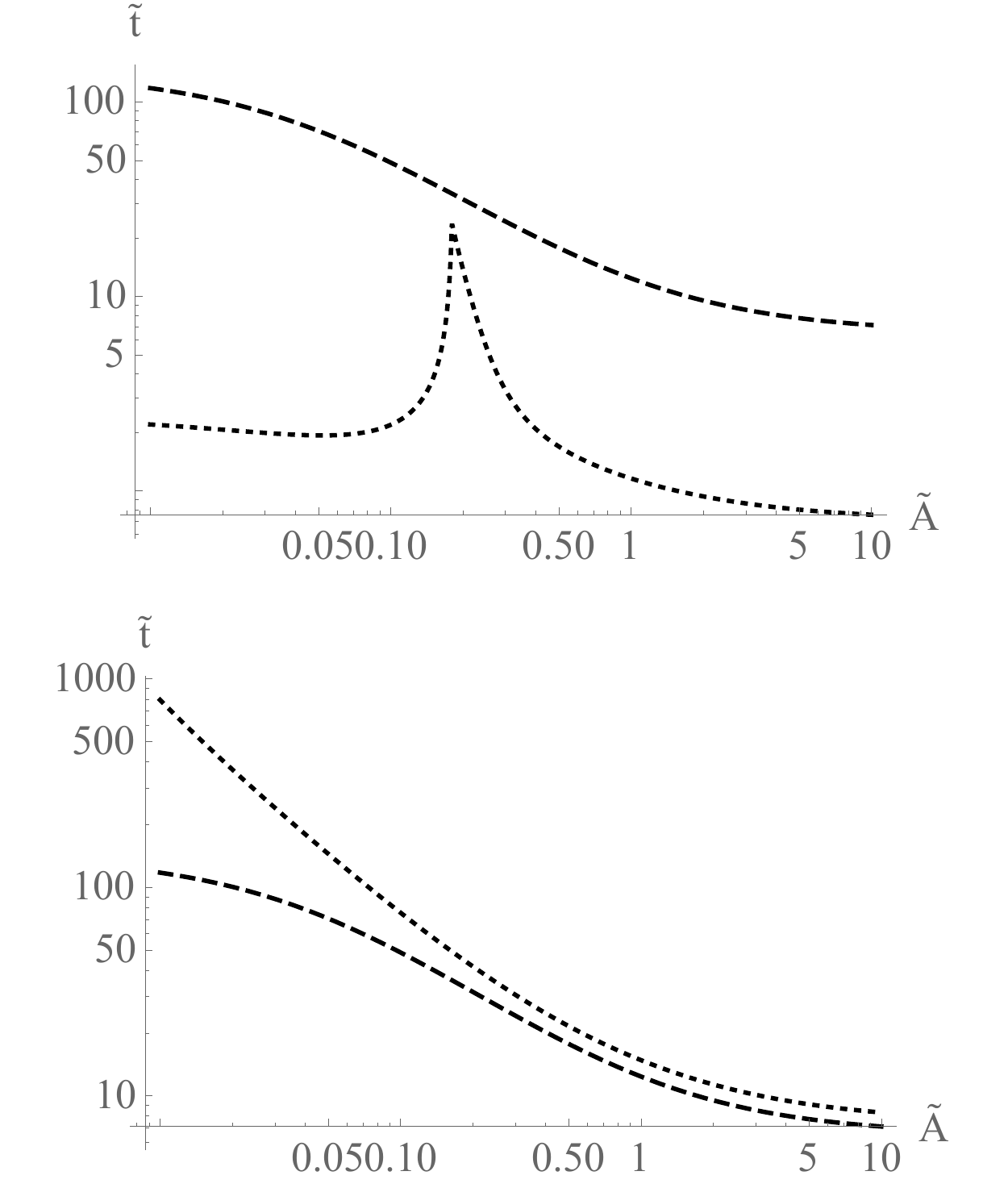}}
	\caption{Timescales of streaming instability (dotted/black) and dust drift (dashed/black) at the outer edge of the pressure bump. Top: inviscid disc, instability grows faster than dust drifts. Bottom: viscous disc ($\alpha = 10^{-3}$), the two timescales are comparable for significant pressure bumps. Parameters correspond to the \texttt{linA} problem.}
	\label{fig:timescales}
\end{figure}
We now verify that the assumptions made in Sect.~\ref{sec:results} are consistent with the global evolution of the disc. The dust-to-gas ratio, the stopping time and the background velocities have been assumed to be constant during the time $t_{\rm growth}$ it takes for the instability to grow. Fig.~\ref{fig:timescales} compares $t_{\rm growth}$ to the drift time of dust into the bump $t_{\rm drift}$ and shows that in the inviscid case, the disc remains in a steady state during the growth of the perturbation since $t\mrm{growth} \ll t\mrm{drift}$. In viscous discs, $t_{\rm growth}$ increases by one order of magnitude and the two timescales become comparable. Long-lived global numerical simulations are therefore required to determine how the instability develops in this case. Moreover, the local depletion of gas caused by back-reaction empties the bump. Its lifetime is compared to the growing time by estimating the ratio $t_{\rm drift} / t_{\rm growth}\epsilon$. As long as $\epsilon \sim 1$, this condition is similar to the precedent one. Hence, our assumptions are essentially valid in real discs and streaming instability is expected to develop linearly as described in Sect.~\ref{sec:results}.

\citet{GLM2017} found that formation of pressure bumps can be triggered by the sole action of dust drag onto the gas, with a spatial resolution insufficient to capture streaming instability. Their Fig.~9 shows that the relative amplitude of the bump is a least of order a few ten per cent, even for viscous discs with $\alpha = 10^{-2}$. Streaming instability in self-induced dust traps offers therefore a promising way to build planetesimals at specific locations. Millimetre grains are converted efficiently into planetesimals -- and potentially later on into planets -- in the trap and only there. Such a scenario is consistent with the dark rings probed recently by ALMA. Similarly, streaming instability may also grow specifically at the edges of the gap created by a massive planet. Since the tidal torque from the planet acts both on dust and gas, it does not induce additional differential velocities and we expect the results derived above to hold. \citet{Johansen2009} found that for large enough metallicities ($Z \gtrsim 0.03$), dust clumps formed by streaming instability collapse gravitationally. We expect a similar trend in a pressure bump, although numerical simulations at high resolution are required to investigate the non-linear stages of the instability.

%----------------------------------------------------------------------------------------
\vspace{-0.5cm}
\section{Conclusion}
\label{sec:conclusion}

In this study, we extend the linear theory of streaming instability in dusty protoplanetary discs to include the eventual presence of local pressure maxima, where solids drift in and pile-up. We find that for pressure bumps with relative amplitude larger than $\sim 20\%$, a bifurcation occurs, giving birth to a novel unstable mode. The instability is powered by the strong differential advection locally imposed by the bump and as such, is resilient against viscous damping from the gas. The growth rate of the instability is typically reduced by one order of magnitude compared to the inviscid case. Hence, in viscous discs, streaming instability is found to grow in and only in local pressure maxima. In particular, streaming instability in self-induced dust traps provides a scenario for the early stages of planet formation consistent with the recent millimetre observations of dark rings in young discs. Numerical simulations are required to investigate the further non-linear development of the instability.

\bibliography{pressure_max}

%%%%%%%%%%%%%%%%%%%%%%%%%%%%%%%
%%%%%%%%%% APPENDICES %%%%%%%%%
%%%%%%%%%%%%%%%%%%%%%%%%%%%%%%%

\appendix

\section{Parameters of the linear tests}
\label{sec:lin}
The sets of parameters for the \texttt{linA}, \texttt{linB}, \texttt{linC} and \texttt{linD} problems are provided in Tab.~\ref{table:param}.
\FloatBarrier
\begin{table}
\centering
\caption{Parameters of the \texttt{linA}, \texttt{linB}, \texttt{linC} and \texttt{linD} problems, where $\eta_0 r_{0} \Omega_{0}/c\mrm{s} = 0.05$.}
\begin{tabular}{c|ccccc}
		Problem 		& $\tilde{k}_x$ & $\tilde{k}_z$ & $\tau\mrm{s}$ & $\epsilon$ 	& $\tilde{s}$ \\
		\hline
		\texttt{linA} 	& 30			& 30			& 0.1			& 3				& 0.4190204 \\
		\texttt{linB}	& 6				& 6				& 0.1			& 0.2			& 0.0154764 \\
		\texttt{linC}	& $1\,500$		& $1\,500$		& 0.01			& 2				& 0.5980690 \\
		\texttt{linD}	& $2\,000$		& $2\,000$		& 0.001			& 2				& 0.3154373
\end{tabular}
\label{table:param}
\end{table}

\section{Solving for background velocities}
\label{sec:vel}
Solving Eqs.~\ref{eq:1} -- \ref{eq:8} by elimination leads to a 256th-order polynomial equation in one of the variables (e.g. $a\mrm{g}$), making the numerical resolution difficult. Simplifications can be made up to almost obtain an analytical solution. Combining Eq.~\ref{eq:1} $\pm$ Eq.~\ref{eq:5}, Eq.~\ref{eq:2} $\pm$ Eq.~\ref{eq:6}, Eq.~\ref{eq:3} $\pm$ Eq.~\ref{eq:7}, Eq.~\ref{eq:4} $\pm$ Eq.~\ref{eq:8}, we obtain
\begin{align}
&a^+ a^- - 2\Omega_0\alpha^- + \gamma^+ a^- + \Gamma\Omega_0^2 = 0\,, \label{eq:I}\\
&\dfrac{1}{2}a^+ b^- + \dfrac{1}{2}a^- b^+ - 2\Omega_0\beta^- +\gamma^+ b^- + 2r_0\eta\Omega_0^2 = 0\,, \label{eq:II}\\
&\dfrac{1}{2}a^+\alpha^- + \dfrac{1}{2}a^-\alpha^+ + 2\Omega_0 a^- + \gamma^+\alpha^- = 0\,, \label{eq:III}\\
&\dfrac{1}{2}b^+\alpha^- + \dfrac{1}{2}b^-\alpha^+ + 2\Omega_0 b^- + \gamma^+\beta^- = 0\,, \label{eq:IV}\\
&\dfrac{1}{2}(a^+)^2 + \dfrac{1}{2}(a^-)^2 - 6\Omega_0^2 - 2\Omega_0\alpha^+ + \gamma^- a^- - \Gamma\Omega_0^2 = 0\,, \label{eq:V}\\
&\dfrac{1}{2}a^+ b^+ + \dfrac{1}{2}a^- b^- - 2\Omega_0\beta^+ + \gamma^- b^- - 2r_0\eta\Omega_0^2 = 0\,, \label{eq:VI}\\
&\dfrac{1}{2}a^+ \alpha^+ + \dfrac{1}{2}a^-\alpha^- + 2\Omega_0 a^+ + \gamma^-\alpha^- = 0\,, \label{eq:VII}\\
&\dfrac{1}{2}b^+\alpha^+ + \dfrac{1}{2}b^-\alpha^- + 2\Omega_0 b^+ + \gamma^-\beta^- = 0\,, \label{eq:VIII}
\end{align}
where
\begin{equation}
\begin{matrix}
a^\pm \equiv a\mrm{p} \pm a\mrm{g}\,, && b^\pm\equiv b\mrm{p} \pm b\mrm{g}\,, \\
\alpha^\pm\equiv\alpha\mrm{p} \pm \alpha\mrm{g}\,, && \beta^\pm\equiv \beta\mrm{p} \pm \beta\mrm{g}\,, \\
\end{matrix}
\end{equation}
and $\gamma^\pm \equiv \dfrac{1\pm\epsilon}{\tstop}$. Eqs.~\ref{eq:I}, \ref{eq:III}, \ref{eq:V} and \ref{eq:VII} contain only $a^\pm$ and $\alpha^\pm$ and can be combined to obtain $a^+$ as a root of a much simpler 5th-order polynomial. No analytic root exists for such a polynomial. However, its single real root can easily be found numerically. From there, the determination of all the other quantities is algebraic, except choosing one root between the two of a second order polynomial. Yet, one of the root is unphysical and can be easily discarded when looking at the numerical values (e.g. $V\mrm{p} \simeq 10^{3} r_{0}\Omega_{0}$). In the case $\eta = 0$ (pure maximum), we obtain analytically $b_{\rm g,p} = \beta_{\rm g,p} = 0$, i.e. the stationary velocities are centred around $x = 0$, as expected. The non-linearity of the system comes from non-linear advection terms, even in the steady state. Although alternative functional forms may solve the equations of motion, we restrain ourselves to a linear form to remain consistent with the shearing box approximation. In absence of any pressure maximum ($\Gamma = 0$), we obtain the expressions given in \citet{Youdin2007}
\begin{align}
&\ol{U}\mrm{g} = \dfrac{2\epsilon\tau\mrm{s}\eta r_0\Omega_0}{\tau\mrm{s}^2 + (1+\epsilon)^2}\,, \\
&\ol{V}\mrm{g} = -\dfrac{3}{2}\Omega_0 x - \eta r_0 \Omega_0 + \dfrac{\epsilon(1+\epsilon)\eta r_0\Omega_0}{\tau\mrm{s}^2 + (1+\epsilon^2)}\,, \\
&\ol{U}\mrm{p} = \dfrac{-2\eta\tau\mrm{s}r_0\Omega_0}{\tau\mrm{s}^2 + (1+\epsilon)^2}\,, \\
&\ol{V}\mrm{p} = -\dfrac{3}{2}\Omega_0 x - \dfrac{(1+\epsilon)\eta r_0\Omega_0}{\tau\mrm{s}^2 + (1+\epsilon)^2}\,.
\end{align}
\label{lastpage}
\end{document}